\documentclass[3p]{elsarticle}
\usepackage{amsfonts}
\usepackage{amsthm}
\usepackage{amsmath}
\usepackage{amscd}
\usepackage{tikz}
\usepackage{verbatim}
\usepackage{indentfirst}
\usepackage{graphicx}
\usepackage{wrapfig}
\usepackage{graphics}
\usepackage{pgfplots,caption}
\usepackage{pict2e}
\usepackage{epic}
\usepackage{geometry}
\usepackage{lineno,hyperref}
\usepackage{multirow}
\setcitestyle{round}

\graphicspath{ {figs/} }
\journal{The International Journal of Clinical Practice}

\begin{document}

\begin{frontmatter}

\title{Predicting Covid-19 EMS Incidents from Daily Hospitalization Trends}


\author{Yangxinyu Xie}
\address{Department of Computer Science, University of Texas at Austin, TX 78712, USA.}
\ead{yx4247@utexas.edu}

\author{David Kulpanowski}
\address{Department of Emergency Medical Services, City of Austin}
\ead{David.kulpanowski@AustinTexas.gov}

\author{Joshua Ong}
\address{Department of Electrical and Computer Engineering, University of Texas at Austin, TX 78712, USA.}
\ead{joshong@utexas.edu}

\author{Evdokia Nikolova}
\address{Department of Electrical and Computer Engineering, University of Texas at Austin, TX 78712, USA.}
\ead{nikolova@austin.utexas.edu}

\author{Ngoc Mai Tran}
\address{Department of Mathematics, University of Texas at Austin, TX 78712, USA.}
\ead{ntran@math.utexas.edu}

\begin{abstract}
{\it Introduction}: The aim of our retrospective study was to quantify the impact of Covid-19 on the temporal distribution of Emergency Medical Services (EMS) demand in Travis County, Austin, Texas, and propose a robust model to forecast Covid-19 EMS incidents.\\
{\it Methods}: We analyzed the temporal distribution of EMS calls in the Austin-Travis County area between January 1st, 2019, and December 31st, 2020. Change point detection was performed to identify critical dates marking changes in EMS call distributions, and time series regression was applied for forecasting Covid-19 EMS incidents.\\
{\it Results}: Two critical dates marked the impact of Covid-19 on the distribution of EMS calls: March 17th, when the daily number of non-pandemic EMS incidents dropped significantly, and May 13th, by which the daily number of EMS calls climbed back to 75\% of the number in pre-Covid-19 time. The new daily count of the hospitalization of Covid-19 patients alone proves a powerful predictor of the number of pandemic EMS calls, with an $r^2$ value equal to 0.85. In particular, for every 2.5 cases where EMS takes a Covid-19 patient to a hospital, one person is admitted.\\
{\it Conclusion}: The mean daily number of non-pandemic EMS demand was significantly less than the period before Covid-19 pandemic. The number of EMS calls for Covid-19 symptoms can be predicted from the daily new hospitalization of Covid-19 patients. These findings may be of interest to EMS departments as they plan for future pandemics, including the ability to predict pandemic-related calls in an effort to adjust a targeted response. 
\end{abstract}


\end{frontmatter}


\section{Introduction}

The ongoing outbreak of the coronavirus disease 2019 (Covid-19) has caused overwhelming disruptions to healthcare systems around the globe \cite{world2020coronavirus, Moghadas9122, Willanm1117, guharoy2021race}, especially the emergency medical services (EMS). Researchers have conducted extensive studies on different aspects of the EMS since the outbreak of Covid-19, including the temporal distribution of EMS demand and emergency department visits \cite{LNM20, BOSERUP20201732, hartnett2020impact, jeffery2020trends, lazzerini2020delayed, thornton2020covid, SaudiArabiaCovid, csan2020effects, wonglaura2020all, jensen2020strategies, lucero2020underutilization, butt2020volume, montagnon2021impact}, pre-hospital patient assessment \cite{fernandez2020covid, yang2020clinical}, medical resource availability and allocation \cite{GIBSON2020resource, hick2020duty, devereaux2020optimizing, schreyer2020emergency}, personnel protective equipment \cite{Murphy707, jalili2020should, ehrlich2020defending}, EMS response practices and strategies \cite{spina2020response, cabanas2020covid, gogapc2020Manage, jensen2020strategies} and  ethical considerations \cite{maguire2020ethics, shadyab2021ethnic}. 

\begin{figure}[h]\centering
\caption[Caption for LOF]{Spatial Distribution of EMS Incidents in 2019-2020\footnotemark. Each light pink dot represents an incident during the 2019-2020 period. The total number of EMS calls in was 246,809. Noticeably, EMS incidents were prevalent along I-35 highway in Austin and were most frequent at downtown Austin.}
\includegraphics[width=0.7\textwidth]{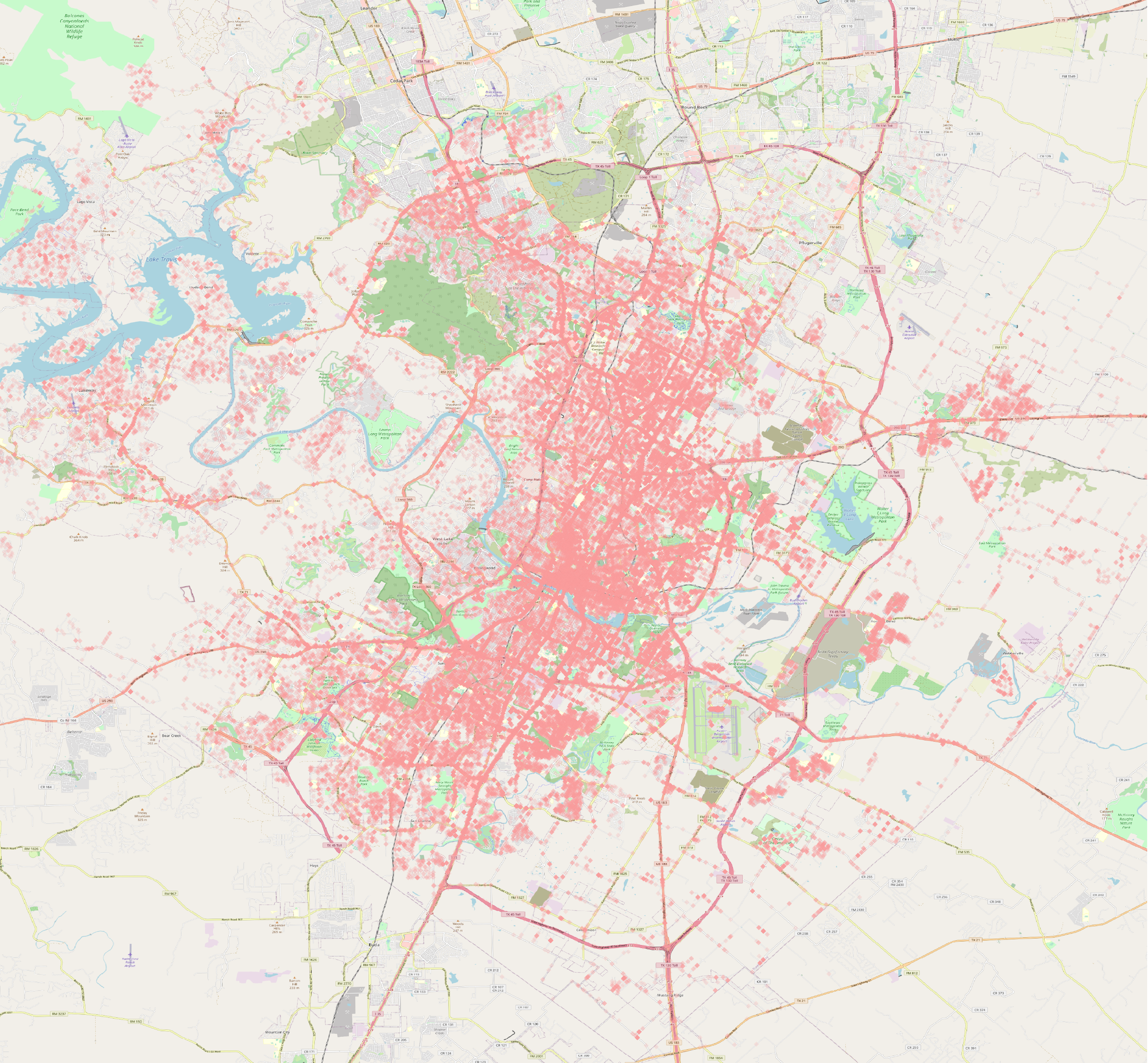}
\label{fig: overall density}
\end{figure}
\footnotetext{This plot is extracted from the City of Austin Open Data Portal: https://data.austintexas.gov/dataset/City-of-Austin-Incidents-Map/m4s3-4qdj}

Globally, the EMS utilization rates varied during the early stage of the Covid-19 outbreak. The number of EMS calls increased in some parts of Europe (France \cite{penverne2020ems}, Ankara, Turkey \cite{csan2020effects}, Copenhagen, Denmark \cite{ jensen2020strategies}) and Saudi Arabia \cite{SaudiArabiaCovid}. However, other observational studies suggested a significant decrease in the number of EMS calls across the United States \cite{LNM20, hartnett2020impact, jeffery2020trends, satty2020ems}, Canada \cite{Ferron21effect}, and other parts of Europe (Italy \cite{lazzerini2020delayed, mantica2020non}, England \cite{thornton2020covid}, Finland \cite{Laukkanen21Early}). Nonetheless, there is very limited information available after the beginning of summer 2020, when states began to lift travel restrictions, and people became better informed of the nature of the virus. 

There is a long history of forecasting that has explored many factors to model EMS demand. The study of forecasting models dates back to the 1970s \cite{hall1971management, aldrich1971analysis}. Earlier studies in cities across the United States incorporated socio-demographic and socioeconomic variables to forecast daily demand \cite{aldrich1971analysis, siler1975predicting, kvaalseth1979statistical, mcconnel1998demand}. Other studies also found that weather factors and seasonality (season, day of the week, and holidays) could also be used to predict the daily ambulance demand \cite{Wong60, mccarthy2008challenge,channouf2007application,setzler2009ems}.

Nevertheless, these EMS models were developed under normal conditions. It is unclear whether they are able to adapt to large-scale disasters such as the Covid-19 pandemic. Studies in the 2016 Melbourne thunderstorm asthma epidemic \cite{andrew2017stormy} discovered a positive correlation between thunderstorm asthma cases and increased EMS demand. To our best knowledge, no models specifically targeting Covid-19 related EMS demand have been proposed. However, earlier studies in France \cite{covid2020early, riou2020emergency} and Israel \cite{JSSZ20} hinted correlations between Covid-19 case hospitalization and Covid-19 EMS demand. 

The objective of this study is twofold. First, we seek to examine the long-term impact of Covid-19 on the temporal distribution of emergency calls and the average response time of ambulance assignment, dispatch and arrival. Second, we seek to design a predictive model to forecast the daily number of Covid-19 pandemic EMS calls.

\section{Methods}
\subsection{Data Description}

Our retrospective study is based on two datasets from the City of Austin Open Data Portal. The first dataset contains all records of EMS incidents in Austin Travis County from January 1st, 2019 to December 31st, 2020 \cite{dataset1}. These EMS incident records do not contain any identifiable information. For each incident, this dataset included its problem type, call disposition, priority number, and the date of the incident. Additionally, it included EMS response times, such as the time elapsed, in minutes, to assign the call, dispatch the ambulance, and for the ambulance to arrive. For a comprehensive list of descriptions of the dataset, see Table \ref{table:records}. This data is collected by the 911 call taker and the ambulance crew. As the caller is interrogated using the Medical Priority Dispatch System (MPDS) protocol \cite{MPDS}, the data is entered into the Computer Aided Dispatch (CAD) software. Once enough information is gathered to generate a unique call, the data is transferred into a Microsoft SQL Server database. The data appears on the Mobile Data Center (MDC) in the ambulance showing the address and nature of the emergency. The ambulance crew then acknowledges receipt of the call and travels en route to the emergency. From this point onward, the ambulance crew enters the information into the MDC. The timestamps gathered by the ambulance crew are the time elapsed, in minutes, to assign the call, dispatch the ambulance, and for the ambulance to arrive.

Please note that in our study, we define incidents of priority number 1 or 2 as high-priority incidents. These incidents are of very high acuity and require a very fast response. Examples include cardiac arrest, severe bleeding, and unconsciousness. For a detailed description of the level of risk corresponding to each priority number, please see Table \ref{table:priority}.

The second dataset consists of the daily frequencies of all incidents from January 1st, 2019 to December 31st, 2020 \cite{dataset2}.This aggregated dataset distinguishes Covid-19 pandemic incidents from incidents of other problem types, as well as the defunct incidents from incidents of other types of call dispositions. The 911 call taker interrogated the caller on breathing difficulty, level of alertness, vomiting, chest pain, chills and sweats, and then, via the MPDS protocol, the call taker decided whether an incident was classified as a Covid-19 pandemic incident. Defunct calls are defined as the EMS incidents where neither the call was from another government agency, nor did the ambulance transport the patient to a hospital. For a complete list of descriptions of the subcategories of defunct calls, see Table \ref{table: disposition}. The Covid-19 hospitalization data consists of the daily count of the hospital admission of Covid-19 patients in the Austin-Round Rock metropolitan statistical area. This hospitalization data was requested from the University of Texas Covid-19 Modeling Consortium \cite{tecaustin}. For a comprehensive list of descriptions of each series of daily frequencies in the dataset, see Table \ref{table:frequencies}.

\begin{figure}[h]\centering
\caption[Caption for LOF]{6 Major Hospitals in Austin\footnotemark. The center of each circle represents the geolocation of each hospital. The size of the circle is proportional to the percentage of EMS incidents disposed to the hospital (see Figure \ref{fig: call disposition}). The two biggest hospitals are Dell Seton Medical Center and South Austin Hospital.}
\centering
\includegraphics[width=0.7\textwidth]{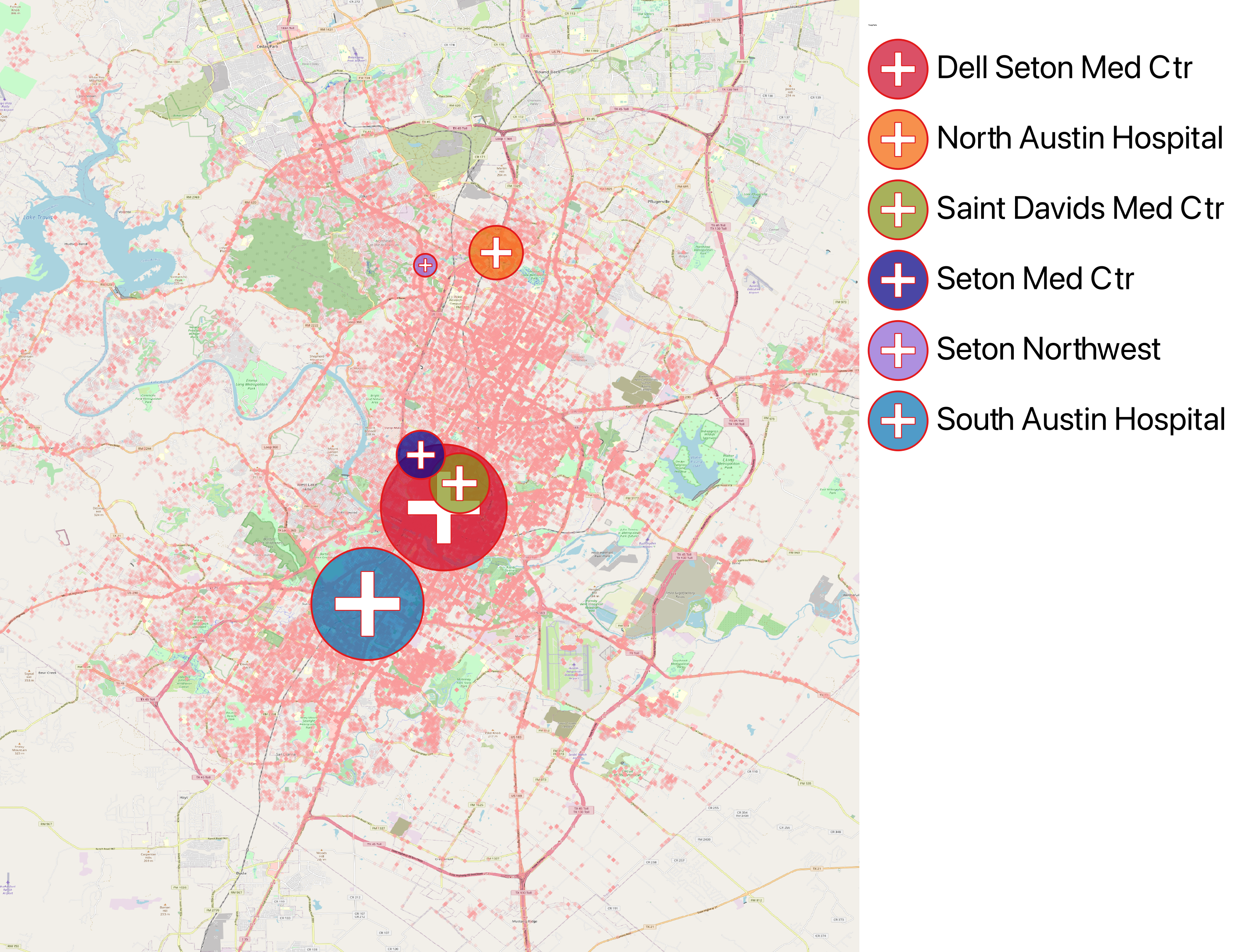}
\label{fig: 6 hospitals}
\end{figure}
\footnotetext{This plot is extracted from the City of Austin Open Data Portal: https://data.austintexas.gov/dataset/City-of-Austin-Hospitals-Map/df29-4hn7}

\begin{table}[h!]
\centering
\begin{tabular}{|p{4cm} | p {3cm} |p{8cm}|} 
 \hline
 Raw names & Names & Description\\
 \hline
 \hline
Problem & Problem Type & The problem type of the emergency. \\
 \hline
Priority\_Number & Priority Number & The acuity of the emergency. A $1$ indicates highest priority and a $15$ would indicate least priority. \\
 \hline
Time\_PhonePickUp\_Date & Date& The date of the EMS call.\\
 \hline
Time\_First\_Unit\_Assigned & Assignment Time &The length of time, in minutes, after the EMS call was picked up and before the first ambulance received notification of the emergency and was assigned.\\
 \hline
Time\_First\_Unit\_Enroute & Dispatch Time & The length of time, in minutes, after the first ambulance was assigned and before the ambulance set off toward the emergency.\\
 \hline
Time\_First\_Unit\_Arrived & Arrival Time & The length of time, in minutes, after the first ambulance wheels began to roll and before the ambulance arrived at the emergency and the wheels stopped.\\
 \hline
Call\_Disposition & Call Disposition & The final disposition of the event, such as cancelled call, transported to hospital, etc. If the ambulance transported the patient to a hospital, the  name of the hospital would be specified. If the emergency call was from another government agency, for example, the Austin Fire Department, then the call disposition would be labelled as "referred". Other types of call dispositions fall into the category of "defunct calls", whose subcategories are defined in Table \ref{table: disposition}; in this case, the call disposition of an incident would be labelled as its respective subcategory. \\
 \hline
\end{tabular}
\caption{ATCEMS database: incident records}
\label{table:records}
\end{table}

\begin{table}[h!]
\centering
\begin{tabular}{|p{5.5cm} | p{4 cm}|p{5.5cm}|} 
 \hline
 Raw names & Names & Description\\
 \hline
 \hline

total\_ts & Total & The daily frequency of all EMS incidents.\\
 \hline
defunct\_calls\_pandemic\_removed\_ts & Non-pandemic Defunct & The daily frequency of non-pandemic defunct EMS incidents. \\
 \hline
Pandemic\_defunct\_calls\_ts & Pandemic Defunct & The daily frequency of pandemic defunct EMS incidents.\\
 \hline
Incidents\_pandemic\_removed\_ts & Non-pandemic Effect & The daily frequency of non-pandemic, non-defunct EMS incidents.\\
 \hline
Pandemic\_effect\_ts & Pandemic Effect & The daily frequency of pandemic, non-defunct EMS incidents.\\
 \hline
hospitalisation\_ts & Hospitalization & The daily frequency of the newly admitted Covid-19 patients to hospitals in the Austin-Round Rock metropolitan statistical area.\\

\hline
\end{tabular}
\caption{ATCEMS database: incident daily frequencies}
\label{table:frequencies}
\end{table}

\begin{table}[h!]
\centering
\begin{tabular}{|p{2cm}|p{3cm}|p{10cm}|} 
 \hline
 Priority & Priority Number & Description\\
 \hline
 \hline
High-priority & 1 & An imminent emergency requiring a very fast response. For example, it can be a person having a cardiac arrest and not breathing, or a person bleeding profusely.\\
 \hline
High-priority & 2 & Incidents also of very high acuity. For example, it can be a person having a cardiac arrest but still breathing, or a person falling off a roof, but breathing and conscious.\\
 \hline
Mid-priority & 3 & Incidents such as a car accident with low level injuries. The patient is breathing, conscious and alert.
\\
 \hline
Mid-priority & 4 & Incidents of a lower level such as a patient who is conscious and stable but still needs transport. For example, it can be a patient who steps on glass and is bleeding, but not profusely and the patient is not in imminent danger.\\
 \hline
Low-priority & 5 & Incidents which could be responded to without lights and sirens. For example, it can be a stubbed toe or a very low level injury.\\
 \hline
Low-priority & 6 to 15 & Incidents of very low priority.\\

\hline
\end{tabular}
\caption{Description of Incident Priority}
\label{table:priority}
\end{table}

\begin{figure}[h]\centering
\caption{Distribution of Priority Number among EMS incidents. 1-2 corresponds to high-priority, 3-4 corresponds to mid-priority and 5-15 corresponds to low-priority.}
\includegraphics[width=0.7\textwidth]{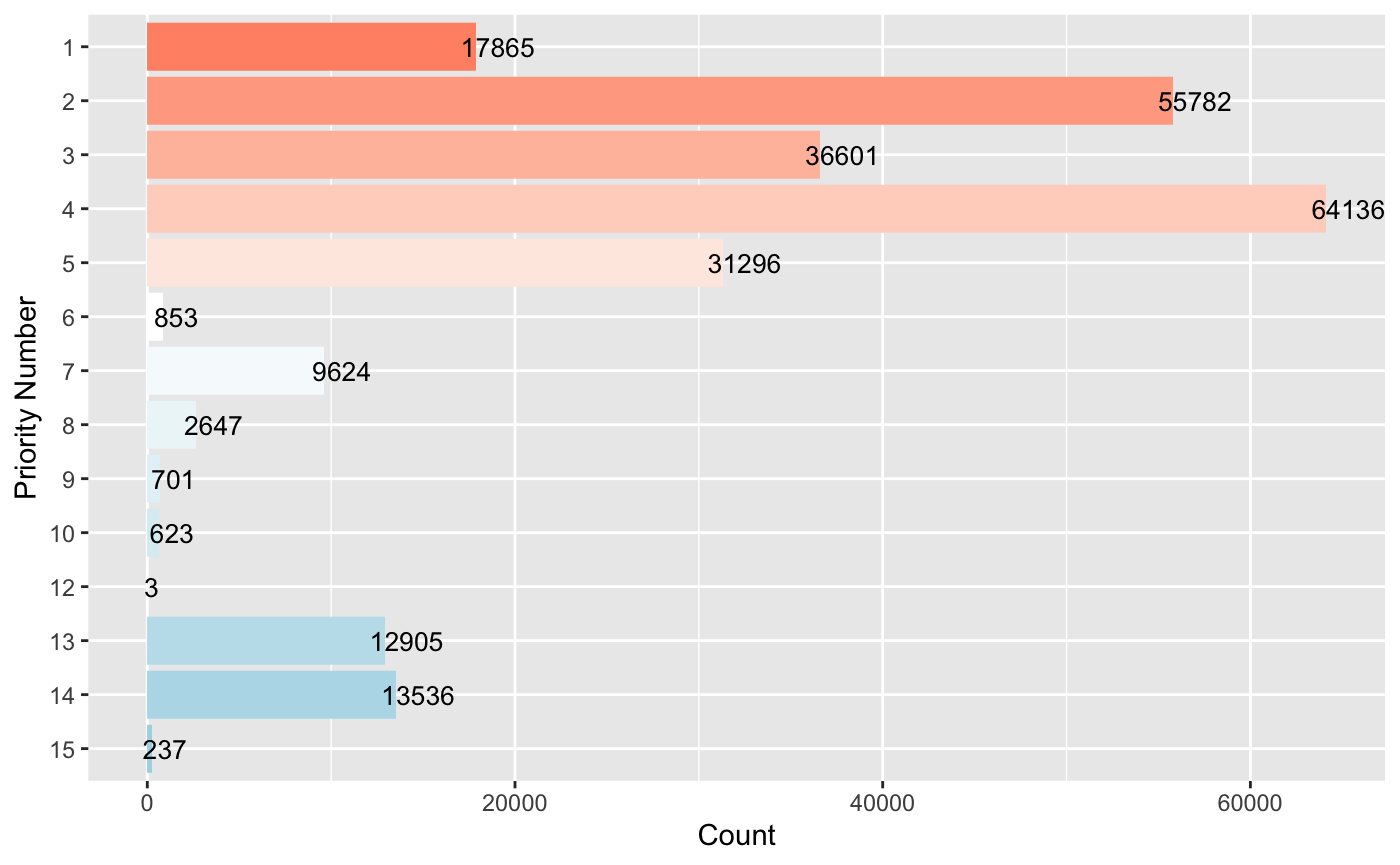}
\label{fig: priority}
\end{figure}

\begin{figure}[h]\centering
\caption{Call Disposition of EMS Incidents in 2019-2020. The percentage was calculated as the number of incidents with a type of disposition against the total number of incidents, which was 246,809. The 6 major hospitals in Travis County are Dell Seton Medical Center, South Austin Hospital, Saint Davids Medical Center, North Austin Hospital, Seton Medical Center and Seton Northwest Hospital. Noticeably, some types of defunct calls had high percentages, including refusal (20\%), cancellation (7\%) and missing patient (5\%).}
\centering
\includegraphics[width=0.7\textwidth]{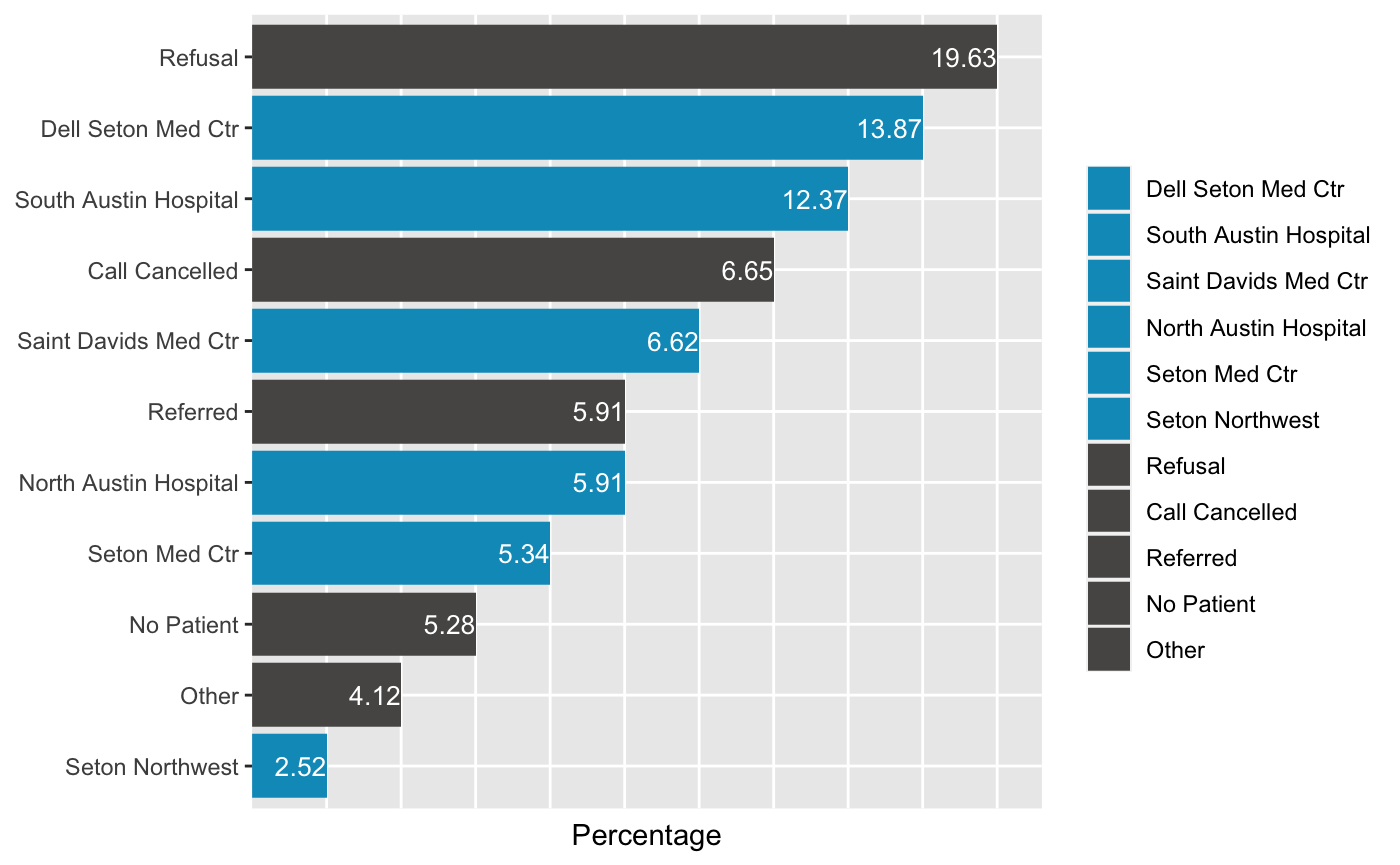}
\label{fig: call disposition}
\end{figure}

\begin{figure}[h]\centering
\caption{Major Problem of EMS Incidents in 2019-2020. There were 22 types of problems with the highest number of calls, which in sum comprise 88\% of the total number of incidents. The frequency of each major problem was above 3000 in the period of 2019-2020. The red color highlights the Covid-19 pandemic. The purple color signifies the problem types with a significant drop in period three in comparison to period one, and the gray color matches with the problem types without a significant drop (see Table \ref{table:decrease} in the result section.)}
\includegraphics[width=0.7\textwidth]{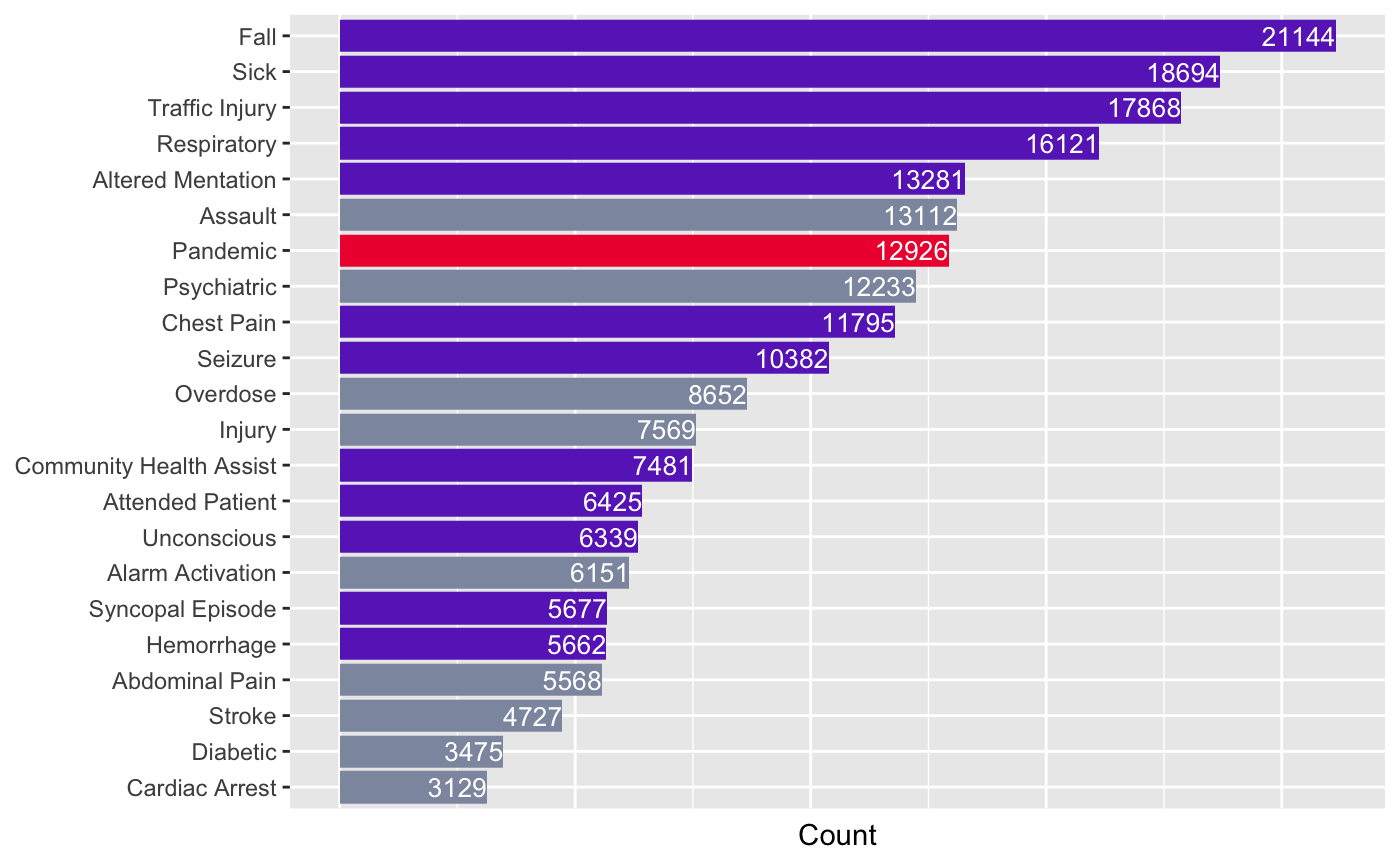}
\label{fig:overall problem}
\end{figure}

\begin{table}[h!]
\centering
\begin{tabular}{|p{4cm} | p{11cm}|} 
 \hline
 Call Disposition & Description\\
 \hline
 \hline
Refusal &	 A refusal means the situation when the ambulance arrives on scene and speaks with the patient, the patient refuses medical help. \\
 \hline
No Patient	 & The ambulance responds to the address provided but does not find a patient.\\
 \hline
Other & Call disposition is unspecified. \\
 \hline
Call Cancelled & \\
 \hline
Duplicate Call & \\
 \hline
False Alarm Call & \\
 \hline
Information Call Only & \\
 \hline
\end{tabular}
\caption{The subcategories of defunct calls}
\label{table: disposition}
\end{table}

\subsection{Statistical Analysis}

The statistical analysis was performed using R statistical computing and graphics environment. The R code is available on GitHub.\footnote{\href{https://github.com/Xieyangxinyu/AustinEMS2020}{https://github.com/Xieyangxinyu/AustinEMS2020})} To evaluate the impact of Covid-19 on non-pandemic incidents, we applied change point detection from the R software package "changepoint.np". We identified changes in mean, and variance (the function "cpt.meanvar") with binary segmentation method \cite{scott1974cluster} and Bayesian information criterion (BIC) penalty assuming underlying normal distribution. To restrict our attention to the impact of the pandemic only, we chose the maximal number of change points as 2. 

After identifying the change point dates, Student's t-tests were applied to make various comparisons between the first period (pre-pandemic) and the third period (post-pandemic). To identify which of the 22 major problems (see figure \ref{fig:overall problem} for a complete list) were impacted by the outbreak of Covid-19 pandemic, we use Student's t-test with Bonferroni correction to compare the mean number of daily calls for each problem type between the first period and the third period. In other words, we set the significance cut-off at $0.05/22 = 0.002273$. Another comparison of the mean time of EMS responses (assignment, dispatch and arrival) was made using one-sided Student's t-test.

To predict the daily frequency of pandemic EMS incidents, time series regression with change point detection was applied in the modeling process. The regressor is the daily frequency of the Covid-19 hospitalization (the "Hospitalization" time series in Table \ref{table:frequencies}). First, we identified multiple change points in the daily Covid-19 hospitalization data. Specifically, we applied exact change point detection, the pruned exact linear time (PELT, \cite{killick2012optimal}) method, on variance with modified Bayesian information criterion (MBIC) penalty assuming underlying normal distribution. Second, to reduce noise and thus improve the predictive power of our model, we smoothed the time series of daily Covid-19 hospitalization by an average of a period of 7 days. Smoothing the daily hospitalization is reasonable because in practical settings epidemiological models commonly output hospitalization forecasts as "expected values," which would then be fed into our model as inputs. Next, we parsed the daily hospitalization data into different periods in accordance with detected change point dates by adding dummy variables. Lastly, after splitting the training (80\%) and testing set (20\%), we applied exhaustive (non-stepwise) selection of autoregressive integrated moving average (ARIMA) models and binomial thinning \cite{Wei08}. The function used was "auto.arima" from the R software package "forecast." To evaluate the robustness of our model, we estimated the $r^2$ value using smoothed daily pandemic EMS incident data to remove unnecessary random noise. To evaluate the predictive power of our model, we estimated the mean squared error and standard error of prediction residual using the original daily pandemic EMS incident data.

\section{Results}

This section consists of three major parts. The first subsection includes the descriptive statistics concerning the big picture of the EMS in Austin. During the 2019 to 2020 period, the total number of EMS calls was 246,809, with the six major hospitals covering 47\% of all EMS incidents. Moreover, the average time it took for a dispatched ambulance to arrive on the scene was 7.14 minutes.

The second subsection identifies two critical dates, March 17th and May 13th, indicating the change in mean and variance of the temporal distribution of non-pandemic EMS calls. (Unless otherwise specified, in the result section, we use non-pandemic EMS calls to refer to non-pandemic effect EMS calls and pandemic EMS calls to refer to pandemic effect EMS calls.) To further investigate the lasting impact of the outbreak of Covid-19, we separated the timeline into three periods: period one, from January 1st, 2019 to March 17th, 2020 with $n = 442$, represents the pre-pandemic norm; period two, from March 18th to May 12th in 2020, represents the sudden outbreak of Covid-19 in Austin; and period three, after May 13th in 2020, represents the new normal. Comparing period one and three, we found that the average daily number of non-pandemic EMS incidents dropped, and the average EMS response time became slower. 

In the third subsection, we propose a time series regression model predicting the daily frequency of pandemic EMS incidents. The regressor, the daily frequency of the Covid-19 hospitalization, is parsed into four stages since March 2020. The regression model obtained an $r^2$ value equal to $0.85$ and indicated that for every 2.5 cases where EMS took a Covid-19 patient to a hospital, one person was admitted.

\subsection{Overall Statistics}

The total number of EMS calls during the 2019 to 2020 period was 246,809. On average, the 37 ambulances in Austin Travis County responded to $338\pm34$ calls per day. Out of all incidents, 29.8\% are of high-priority, 40.8\% are of mid-priority, and 29.3\% are of low-priority, as shown in Figure \ref{fig: priority}. The spatial distribution of EMS incidents is shown in Figure \ref{fig: overall density}. Noticeably, EMS incidents were prevalent along I-35 highway in Austin and were most frequent in downtown Austin. There are six major hospitals in Travis County (for the locations of these hospitals, see Figure \ref{fig: 6 hospitals}; for the percentages of EMS incidents classified by call dispositions, see Figure \ref{fig: call disposition}), among which Dell Seton Medical Center and South Austin Hospital are the two biggest ones, covering 26\% of all EMS incidents. In summary, the six major hospitals took care of 47\% of all EMS incidents. Meanwhile, refusal (20\%), cancellation (7\%), missing patient (5\%) are also among the highest percentages.

There were 22 types of problems with the highest number of calls, which in sum comprise 88\% of the total number of incidents. For a complete list of problem types and their frequencies in the period of 2019 to 2020, see Figure \ref{fig:overall problem}. Notably, there were 12,926 emergency calls related to the Covid-19 pandemic. 

\begin{figure}[h]\centering
\caption{Performance of EMS Responses in 2019-2020}
\includegraphics[width=0.7\textwidth]{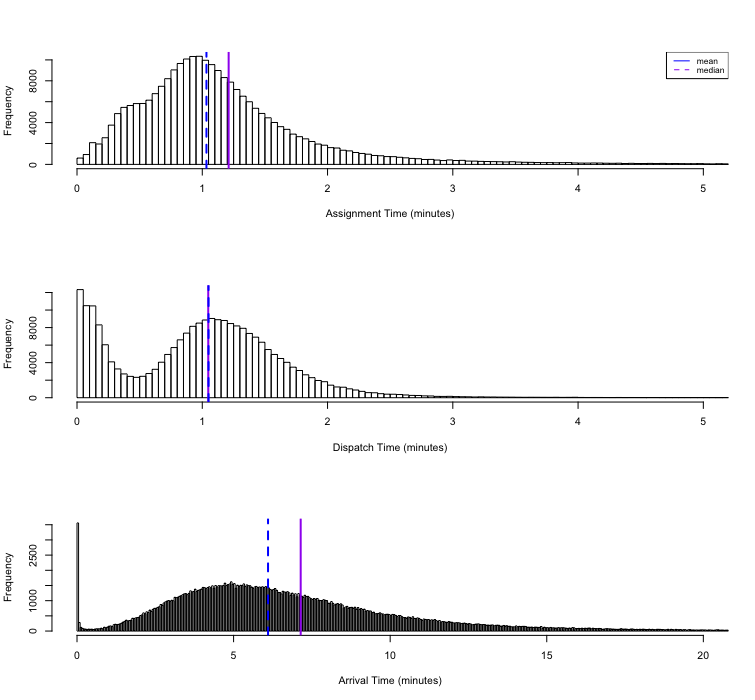}
\label{fig:response time}
\end{figure}

The distribution of EMS response times in 2019-2020 is shown in Figure \ref{fig:response time}. On average, after an EMS call was picked up, the mean time for an ambulance assignment was 1.21 minutes, and the median was 1.03 minutes. The mean and median time for an assigned ambulance to dispatch were both 1.05 minutes. The average time for a dispatched ambulance to arrive on the scene was 7.14 minutes, and the median was 6.1 minutes.\footnote{The average response time varies among the six major hospitals. For a detailed comparison, see the appendix section.}

\subsection{Impacts of the pandemic}

\subsubsection{EMS call volumes changed significantly with two critical dates}

\begin{figure}[h]\centering
\caption{Daily Frequency of EMS Incidents in 2019-2020. The two purple lines mark the two critical change point dates, March 17th and May 13th. Each colored line indicates the temporal distribution of a time series defined in Table \ref{table:frequencies}. The darker color represents the smoothed, 7-day average, trend, while the lighter color represents the raw frequencies. In summary, the total number of non-pandemic EMS calls decreased during Covid-19 times.}
\includegraphics[width=0.7\textwidth]{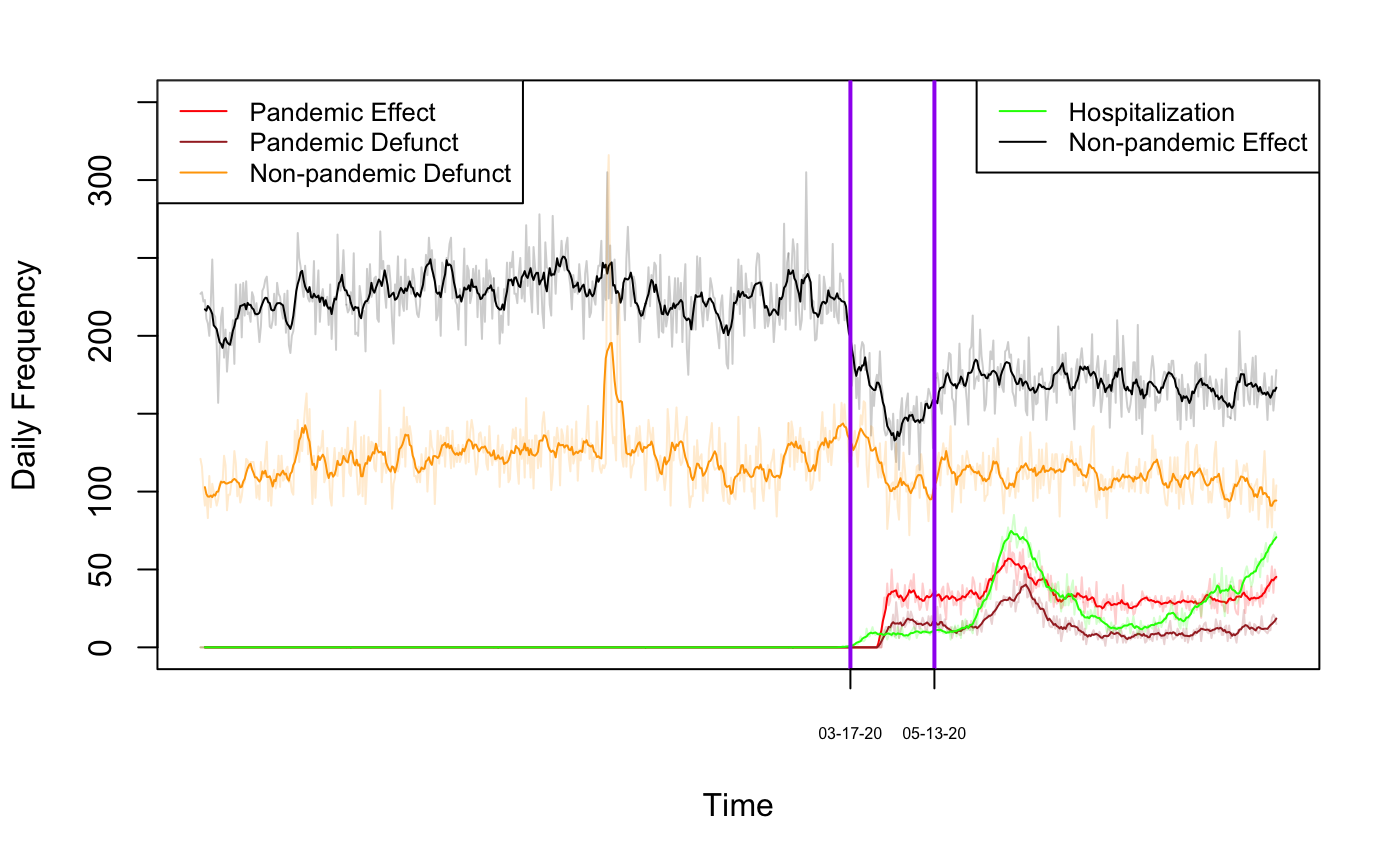}
\label{fig: ts overall}
\end{figure}

Via change point detection in mean and variance of the temporal distribution of non-pandemic EMS calls, with binary segmentation method and BIC penalty, we found two critical dates marking the impact of Covid-19 on non-pandemic EMS calls (shown as purple vertical lines in Figure \ref{fig: ts overall}). The first date was March 17th, around the time of the Austin shelter-in-place order. The average daily number of non-pandemic EMS incidents dropped significantly, from 225.69 in period one to 155.84 in period two. 

The second critical date was May 13th, or the beginning of summer, by which the daily number of EMS calls climbed back to a new plateau, which is about 75\% of that before March 17th (169.53 in period three versus 225.69 in period one). In sum, the total number of non-pandemic EMS calls decreased during Covid-19 times. 

Noticeably, despite the disruption of Covid-19, non-pandemic defunct calls did not witness the same level of decrease in daily frequency (shown by the orange line in Figure \ref{fig: ts overall}). Nevertheless, the proportion of defunct calls rose from 35\% of the overall number of non-pandemic EMS calls in period one to 39\% in period three. For a complete comparison of the number of non-pandemic EMS incidents per day among periods one, two, and three, see Table \ref{table: Comparison Non-Pandemic}.

\begin{table}[h]
\centering
\begin{tabular}{||c c c c c c c||} 
\hline
\multicolumn{7}{||c||}{Comparison of number of non-pandemic EMS incidents}\\
 \hline
 & \multicolumn{2}{c}{period 1} & \multicolumn{2}{c}{period 2} & \multicolumn{2}{c||}{period 3}\\ [0.5ex] 
  & \multicolumn{2}{c}{$n = 442$} & \multicolumn{2}{c}{$n = 56$} & \multicolumn{2}{c||}{$n = 233$}\\ [0.5ex] 
  & Mean & SD & Mean & SD & Mean & SD\\ [0.5ex] 
 \hline\hline
Admitted & 225.69 & 19.43 &  155.84 & 20.23& 169.53 &14.76 \\ 
Defunct & 122.35  & 21.71 & 113.59  & 19.30& 109.34 &12.73 \\
 \hline
\end{tabular}
\caption{Comparison of number of non-pandemic EMS incidents per day among period one (before March 17th, pre-pandemic), period two (March 18th - May 12th), period three (after May 13th, post-pandemic). SD stands for standard deviation. Overall, the total number of non-pandemic EMS calls decreased during Covid-19 times.}
\label{table: Comparison Non-Pandemic}
\end{table}

As for pandemic-related incidents, the red line represents pandemic EMS calls disposed to hospitals, and the brown line represents pandemic related defunct calls in Figure \ref{fig: ts overall}. Note that pandemic EMS calls are closely correlated with the daily frequency of the newly admitted Covid-19 patients to hospitals in the Austin-Round Rock metropolitan statistical area, shown by the green line. Our time-series model in the next subsection found that this hospitalization trend alone is a powerful predictor for pandemic EMS calls. 

\begin{table}[h!]
\centering
\begin{tabular}{||c c c c||} 
\hline
\multicolumn{4}{||c||}{Comparison of number of non-pandemic EMS incidents by problem type}\\
 \hline
 & period 1 & period 3 & \\ [0.5ex] 
  & $n = 442$ & $n = 233$ & \\ [0.5ex] 
  & Mean& Mean & p-value\\ [0.5ex] 
 \hline\hline
 Cardiac Arrest & 4.03 & 4.54& 0.999\\
 Psychiatric & 16.52 & 17.49& 0.996\\
  Assault & 17.86 & 18.45& 0.904\\
 Stroke & 6.51 & 6.55 & 0.561\\
 Abdominal Pain & 7.65 & 7.51& 0.271\\
 Alarm Activation & 8.52 & 8.23& 0.128\\
 Injury & 10.67 & 10.24& 0.061\\
 Overdose & 12.16 & 11.42 & 0.008\\
 Diabetic & 4.91 & 4.45 & 0.007\\
 \hline\hline
 Altered Mentation & 20.90 & 14.15 & $< 0.0001$\\
 Attended Patient & 10.43 & 6.63 & $< 0.0001$\\
 Chest Pain & 19.17 & 11.10 & $< 0.0001$\\
 Community Health Assist & 11.56 & 8.06 & $< 0.0001$\\
 Fall & 30.47 & 27.27 & $< 0.0001$\\
 Hemorrhage & 8.51 & 6.48 & $< 0.0001$\\
 Respiratory & 24.83 & 16.19 & $< 0.0001$\\
 Seizure & 15.18 & 12.91 & $< 0.0001$\\
 Sick & 31.78 & 14.79 & $< 0.0001$\\
 Syncopal Episode & 8.66 & 6.62 & $< 0.0001$\\
 Traffic Injury & 27.39 & 21.37 & $< 0.0001$\\
 Unconscious & 9.46 & 7.68 & $< 0.0001$\\
 \hline
\end{tabular}
\caption{One-sided (greater) student-t comparison of the number of non-pandemic EMS incidents per day among period one (before March 17th), period two (March 18th - May 12th), period three (after May 13th); via Bonferroni correction, $\alpha = 0.00227$}
\label{table:decrease}
\end{table}

\begin{table}[h]
\centering
\begin{tabular}{||c c c c c c c||} 
 \hline
 & \multicolumn{2}{c}{Assignment Time} & \multicolumn{2}{c}{Dispatch Time} & \multicolumn{2}{c||}{Arrival Time}\\ [0.5ex] 
  & Mean & Median & Mean & Median & Mean & Median\\ [0.5ex] 
 \hline\hline
period 1 & 1.18 & 1.02 & 1.01 & 1.00 & 6.88 &5.87 \\ 
period 2 & 1.33 & 1.12 & 1.08 & 1.10 & 7.52 &6.40 \\
period 3 & 1.25 & 1.05 & 1.11 & 1.10 & 7.61 &6.55 \\
 \hline
\end{tabular}
\caption{Comparison of the average EMS response time (in minutes) among period one (before March 17th), period two (March 18th - May 12th), period three (after May 13th). Assignment time, dispatch time, and arrival time were all slower during periods two and three than those in period one.}
\label{table: covid response time}
\end{table}

\begin{figure}[h]\centering
\caption{Distribution of EMS response times (in minutes) among period one (before March 17th), period two (March 18th - May 12th), period three (after May 13th). The mean is represented by the solid line and median by the dashed line. Both mean and median of the three types of response time have slightly shifted to the right from period one to period three.}
\includegraphics[width=0.7\textwidth]{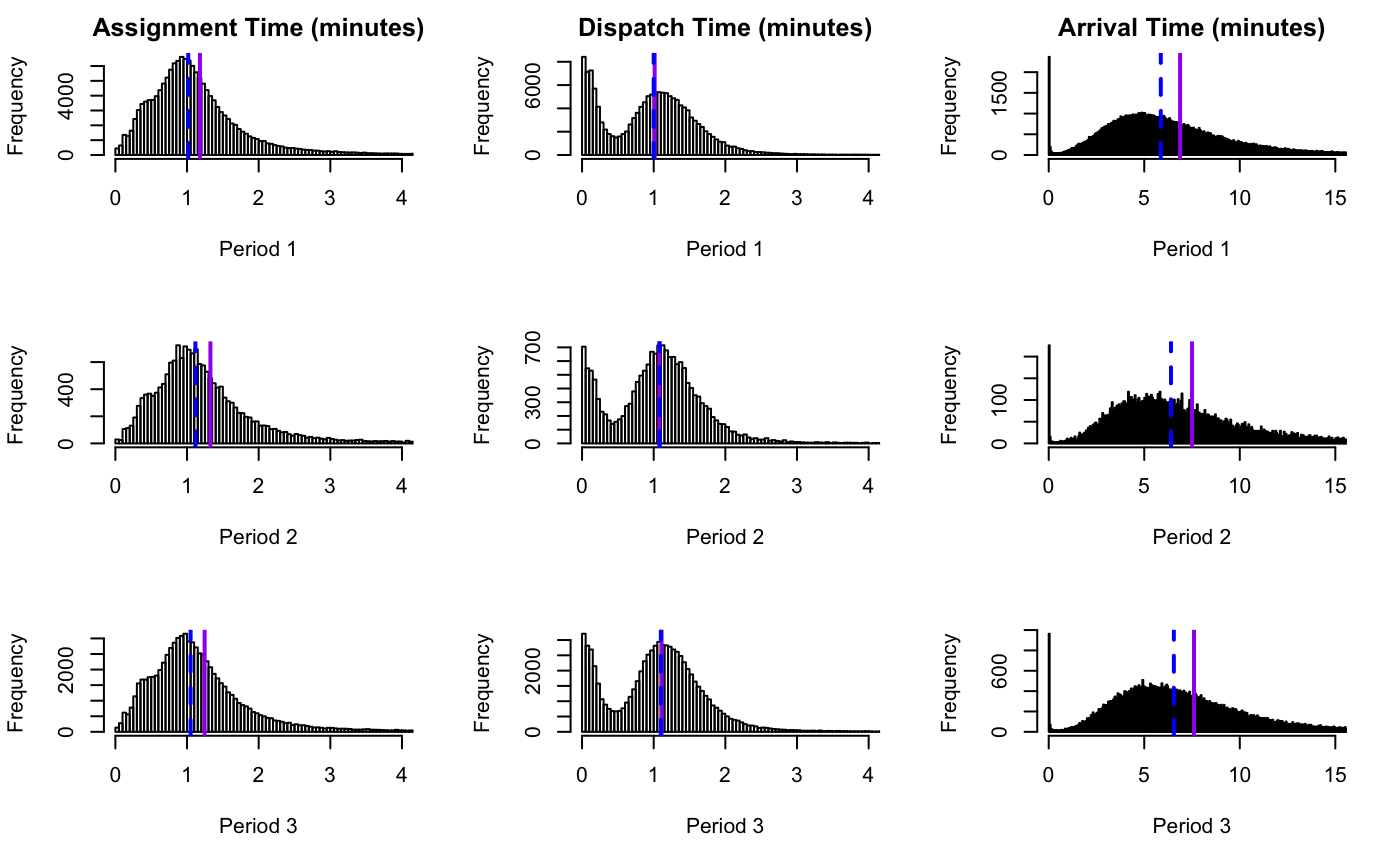}
\label{fig: covid response time}
\end{figure}

\begin{table}[h]
\centering
\begin{tabular}{||c | c c c||} 
 \hline
High-priority Incidents & Period 1 & Period 2 & Period 3\\ [0.5ex] 
 \hline\hline
Total Count & $44790$ & $4823$ & $20402$\\ 
$\le 5$ min & $20023$ & $1671$ & $6854$\\
Proportion & $44.7\%$ & $34.6\%$ & $33.6\%$\\
 \hline
\end{tabular}
\caption{Comparison of the proportions of high-priority incidents EMS whose response time (in minutes) were within 5 minutes among period one (before March 17th), period two (March 18th - May 12th), period three (after May 13th). Assignment time, dispatch time, and arrival time were all slower during periods two and three than those in period one.}
\label{table: covid less 5 min response}
\end{table}

A closer comparison between period one and period three showed that the frequencies of only certain types of problems dropped, while others remained unaffected by Covid-19. The complete list of comparisons of the number of non-pandemic EMS incidents of each problem type can be found in Table \ref{table:decrease}.

\subsubsection{EMS response times were slower during the pandemic}
\label{sec: prolongation of EMS response time}

The performance of EMS response was also impacted since the outbreak of Covid-19. The mean response time for each action, assignment, dispatch and arrival, was slower during period two and period three than that in period one (one-sided t-test, p-value $< 0.01$). In particular, the arrival time during period three is 0.73 minutes slower than during period one for each EMS incident. A complete list of comparisons of the mean and median of response time is in Table \ref{table: covid response time}. The reader can also see from Figure \ref{fig: covid response time} that the mean (solid line) and median (dashed line) of the distributions of the response time has slightly shifted to the right from period one to period three. As shown in Table \ref{table: covid less 5 min response}, the proportions of high-priority incidents EMS whose response time (in minutes) were within 5 minutes decreased from 44.7\% in period one to 33.6\% in period three. A one-sided t-test suggests that this decrease is statistically significant with p-value $< 0.01$.

\subsection{Covid-19 hospitalization predicts pandemic EMS calls}

\begin{figure}[h]\centering
\caption{Covid-19 daily hospitalization. The four change points on variance in the temporal distribution of Covid-19 hospitalization are shown by the purple vertical lines.}
\includegraphics[width=0.7\textwidth]{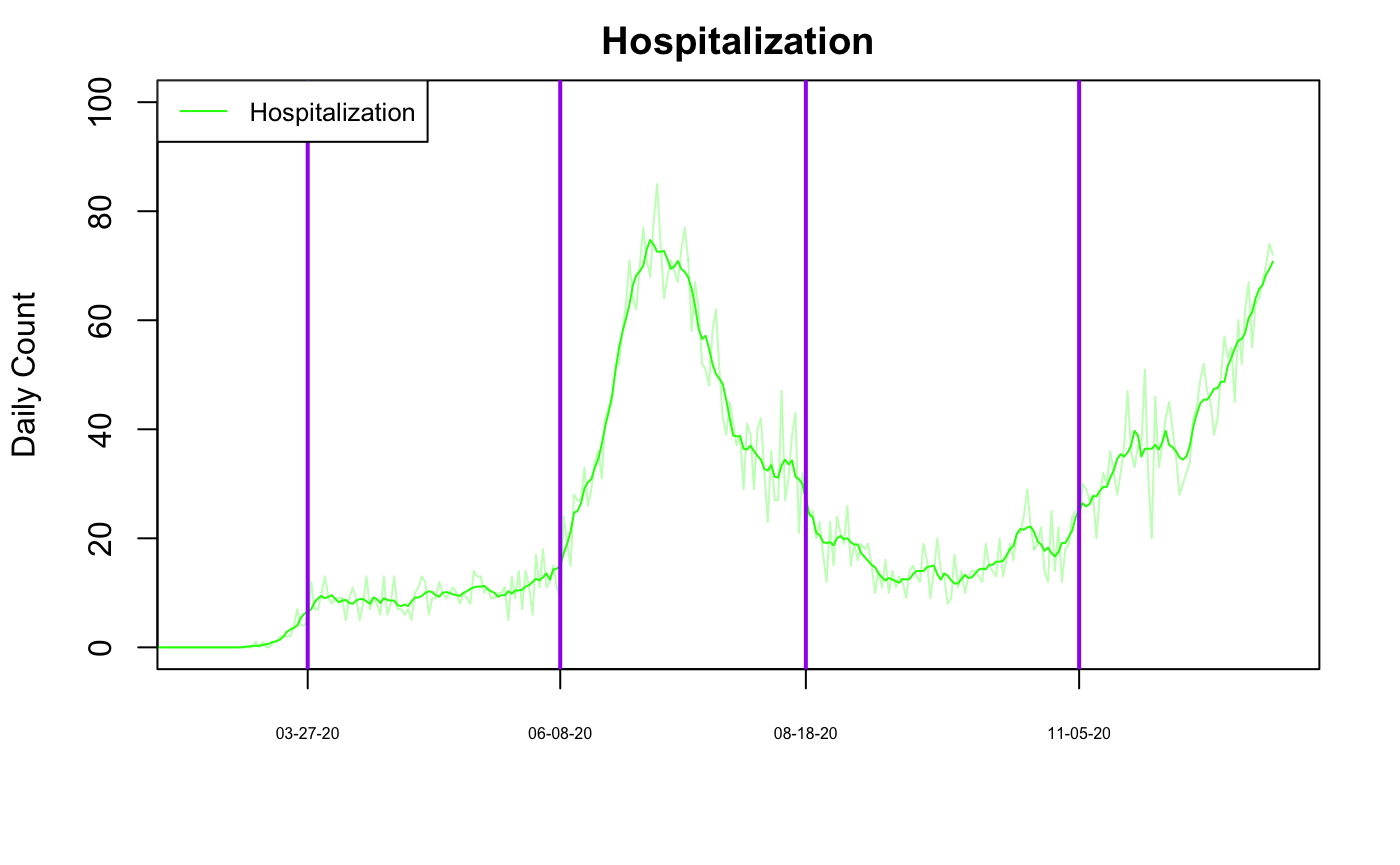}
\label{fig:hospitalization cp}
\end{figure}

\begin{figure}[h]\centering
\caption{Time series regression with change point detection. The darker blue line represents the predicted value; the lighter blue area represents the 95\% prediction interval. The red line represents the actual daily frequencies of pandemic EMS incidents, consistent with Figure \ref{fig: ts overall}.}
\includegraphics[width=0.7\textwidth]{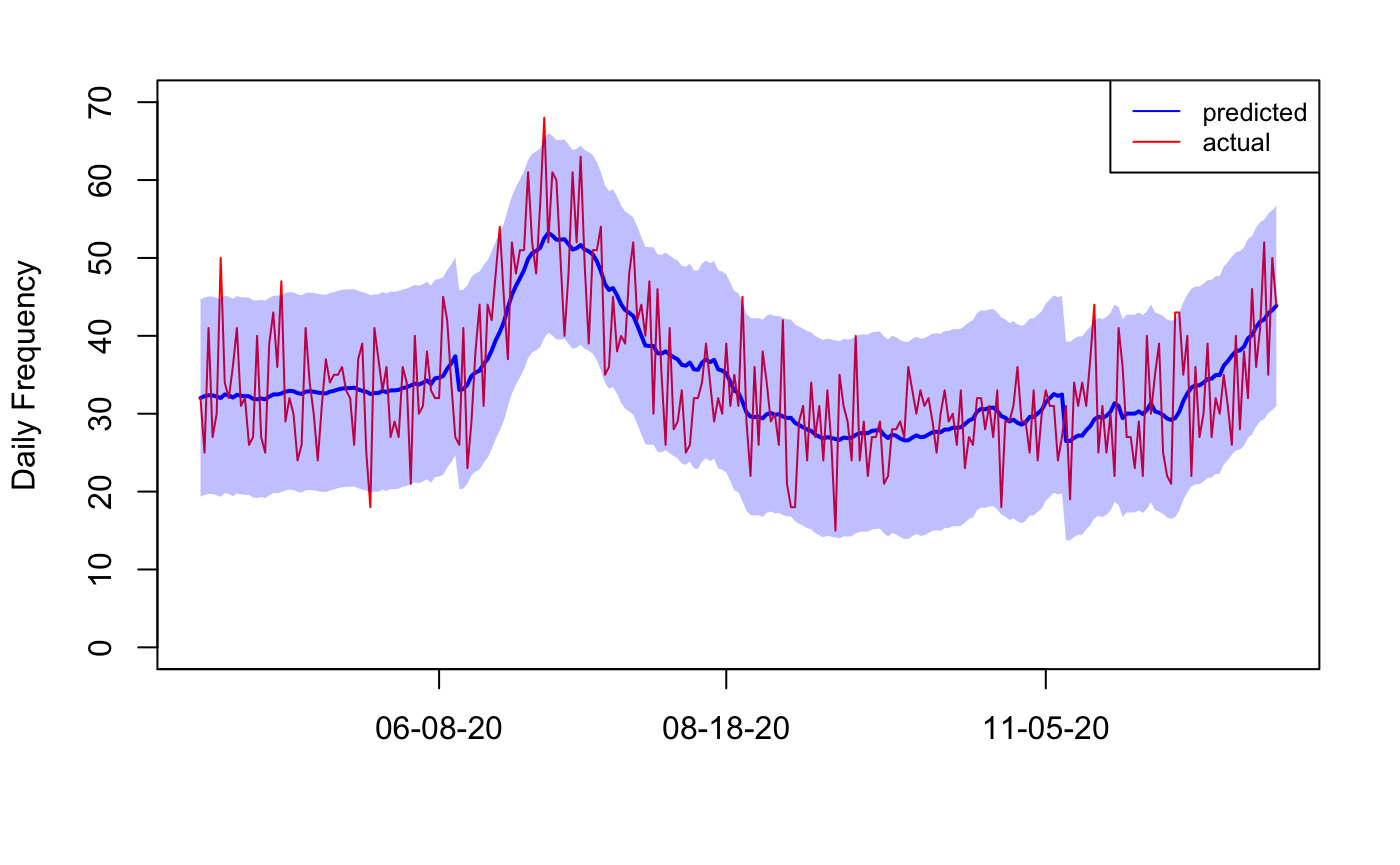}
\label{fig: arima model ts}
\end{figure}

The exact change point detection on variance with MBIC penalty yielded 4 Covid-19 hospitalization stages since March 2020. The four change point dates are shown by the purple vertical lines in Figure \ref{fig:hospitalization cp}. In particular, we see rises in Covid-19 hospitalizations during summer 2020 (June 8th - August 18th) and winter 2020 (November 5th - December 31st). This resonated with the global pandemic second and third waves. To achieve optimal modeling performance, we avoided the uncertainties at the earliest stage of the pandemic by only fitting and testing the model using data from April 9th to December 31st in 2020. Thus, we set the dummy variable "1st\_period" to 1 during the period from April 9th - June 8th, the dummy variable "2nd\_period" to 1 during the period from June 8th - August 17th, and the dummy variable "3rd\_period" to 1 during the period from August 17th - December 31st.

The exhaustive selection of ARIMA models outputs 0 autoregressive order, 0 degrees of differencing, and 0 moving average order. The coefficients of the linear regression model, with the pandemic EMS Calls regressed on the smoothed (7-day average) hospitalization data, are specified in Table \ref{table:coef}. This model obtained an $r^2$ value equal to $0.85$. Moreover, the residual standard error of the training and test set are 6.62 and 5.44, respectively, which mainly characterize the random fluctuations on daily counts. The reader can also see from Figure \ref{fig: arima model ts} that most of the actual daily frequencies of pandemic EMS incidents fall within the 95\% prediction interval. Therefore, daily new hospitalization of Covid-19 cases alone proves a powerful and robust indicator of pandemic-related EMS demand. The 0.40 estimate of slope suggests that on average, for every 2.5 cases where EMS took a Covid-19 patient to a hospital, one person was admitted. Moreover, moving from the 1st to the 3rd period, there was a decreasing offset between pandemic EMS incidents and hospitalization numbers.

\begin{table}[h!]
\centering
\begin{tabular}{||c c c c c c||} 
 \hline
 Coefficients & intercept & hospitalization & 1st\_period & 2nd\_period & 3rd\_period\\ [0.5ex] 
 \hline\hline
Estimate & 15.09774 & 0.40327 &  13.87507 & 7.90718 & 6.72668\\ 
Standard Error & 2.15766  & 0.04341 & 1.99988  & 1.34397 & 1.76075 \\
 \hline\hline
 train & \multicolumn{5}{c||}{$r^2 = 0.85$} \\
 & \multicolumn{5}{c||}{Residual standard error: 6.619 on 209 degrees of freedom}\\
 test & \multicolumn{5}{c||}{$r^2 = 0.84$}\\
 & \multicolumn{5}{c||}{Mean squared error: 29.012 } \\
 & \multicolumn{5}{c||}{Standard error of prediction residual: 5.435}\\
\hline
\end{tabular}
\caption{Coefficients of Time Series Regression with Change Point detection}
\label{table:coef}
\end{table}

\section{Discussion}

The daily number of non-pandemic EMS incidents dropped precipitously since the beginning of the Covid-19 pandemic and the declaration of the local state of emergency. This resonates with earlier studies in the United States \cite{LNM20, BOSERUP20201732, jeffery2020trends, satty2020ems}, Italy \cite{lazzerini2020delayed, mantica2020non}, England \cite{thornton2020covid}, Canada \cite{Ferron21effect} and Finland \cite{Laukkanen21Early}. Even though the number of non-pandemic EMS incidents increased at the beginning of summer, it remained consistently lower than in the pre-pandemic era. More interestingly, the proportion of non-pandemic defunct calls rose after the outbreak of the Covid-19 pandemic. The rise of refusals to transportation to hospitals has also been observed in Israel \cite{siman2021assessment}. This corroborates with earlier studies in the United States suggesting that people showed reluctance to receive hospitalization out of fear of contagion \cite{wonglaura2020all, moroni2020collateral}. 

This significant decrease in non-pandemic EMS incidents may also be partially explained by travel restrictions (attended patient, community health assist, fall, hemorrhage, traffic injury) and misclassification into Covid-19 cases (chest pain, respiratory, and sick). An overall decrease in traffic-related incidents was also observed in Israel \cite{siman2021assessment} and Ontario, Canada \cite{Ferron21effect} during period two. People may have traveled less frequently due to the shelter-in-place order issued in March 2020 and possibly been less exposed to injury-prone locations. A study in Connecticut found that the mean daily vehicle miles traveled and the mean daily counts of crashes significantly decreased in the post-stay-at-home period in 2020 \cite{doucette2021initial}. Our comparison between periods one and three in Table \ref{table:decrease} showed that the frequencies of traffic-related incidents remained lower in the long run than those of the pre-pandemic time, despite the lifting of Texas stay-at-home order. 

The decrease in emergency calls regarding chest pain, sickness, and respiratory problems was also found in Netherlands \cite{Koning21Emergency} and Ontario, Canada \cite{Ferron21effect}. One possible explanation is that calls previously classified as "chest pain," "respiratory," and "sick," due to their similarity to Covid-19 symptoms, became categorized as "pandemic" \cite{Ferron21effect}.

Prior studies have found increase in EMS calls regarding cardiac arrest \cite{siman2021assessment, Ferron21effect, goldberg2021impact}, stroke \cite{Ferron21effect} mental health \cite{siman2021assessment} and overdose \cite{Ferron21effect}, during the early stages of the Covid-19 pandemic. Our study confirmed that, in the long term, the EMS demand for these problems did not decrease. Interestingly, although earlier findings in Ontario, Canada \cite{Muldoon2021violence} showed a decrease in emergency calls related to assault during the period March-May 2020, our study found that the daily average number of emergency calls related to assault in Austin was higher during the post-pandemic time (period three) than pre-pandemic time (period one). 

Despite the overall decrease in emergency call volume, we observed statistically significant prolongation of the EMS response time (see Section \ref{sec: prolongation of EMS response time}). Whether this is clinically significant is beyond our scope, though there are studies that may raise concerns. \cite{wilde2013emergency}. In particular, it was shown that for incidents of intermediate or high risk of mortality, a survival benefit was identified when the response time was within 4 to 5 minutes for patients when compared with that exceeding 5 minutes \cite{blackwell2002response, pons2005paramedic}. Hence, as found in this study, the $11.1\%$ decrease in the proportion of high-priority incidents whose response time was within 5 minutes may raise concerns for EMS practitioners. 

Investigating the cause of the lengthened response time is a very interesting problem, but it is well beyond the scope of this study. One possible explanation is that the adaptation to Covid-19 personal protective equipment (PPE) protocol \cite{centers2020interim, ribeiro2020international, benitez2020adapting} may have contributed to the overall prolongation of EMS response time. Prior studies reported that following Covid-19 PPE protocols has hampered surgical performance by causing visual impairment, communication impediments, and increased fatigue \cite{benitez2020impact}. We hypothesize that the same could have happened to health workers at the EMS department. Additionally, previous studies have reported shortages in resources such as PPE equipment among EMS facilities \cite{ventura2020emergency}, which may limit the number of available personnel and ambulances. Indeed, local news reported in Austin suggested that the EMS department did not receive adequate funding to maintain sufficient personnel and ambulances during the pandemic, leading to delays in response time \cite{Weber20Austin, Tristan20Austin}. Meanwhile, the prolongation of EMS response time after the beginning of the Covid-19 pandemic was also found in Finland \cite{Laukkanen21Early} and Western Pennsylvania \cite{satty2020ems}. 
 
Previous studies reporting the patterns of EMS demand under the impact of Covid-19 in France \cite{covid2020early, riou2020emergency} and Israel \cite{JSSZ20} suggested that the increase of EMS calls for Covid-19 symptoms followed the same shape as for confirmed Covid-19 patients. Our study confirms and quantifies this observation. We computed the correlation between the daily new hospitalization of Covid-19 patients and pandemic-related EMS demand. Covid-19 hospitalization projection models have been investigated extensively \cite{tecaustin, moghadas2020projecting, du2020covid, gel2020covid, ferstad2020model}. Our regression model, which can use forecasts of the Covid-19 hospitalization count as input, can thus serve as a simple and convenient tool for EMS departments to quickly forecast pandemic-related EMS demand. Interestingly, as time progressed, the offset between pandemic EMS and hospitalization numbers became smaller. The reason remains unclear. We hypothesize that people became better at making informed decisions about Covid-19-related EMS Calls. At the beginning of the outbreak, due to a lack of knowledge of the epidemic, people might have made haste to raise red flags even though they were not contaminated by Covid-19. However, we did not find supporting evidence for our hypothesis.

\section{Limitations and Future Directions}

Our analysis and the proposed time series model were restricted to the data observed in Austin. It will be interesting to see whether a similar simple model can be fitted to forecast pandemic-related EMS demand in other cities. Moreover, the decrease in the offset between pandemic EMS and hospitalization numbers stood out very interesting. However, we are unable to find a satisfying reason, and we hope for future studies to test our hypothesis. 

Another limitation was that our regression model forecasts the overall number of Covid-19 EMS cases without differentiating the priority levels or the geolocation of each incident. The ability to identify and forecast urgent incidents may improve EMS further. Additionally, a more fine-grained model incorporating spatial information may help identify incident hotspots within the city.

\section{Conclusions}

This study analyzed the impact of the Covid-19 pandemic on EMS call distributions and response time. Overall, we have confirmed that non-pandemic EMS calls in Austin have significantly decreased since the beginning of the Covid-19 pandemic, yet the proportion of non-pandemic defunct calls increased. In terms of problem types, we observed that some problems seemed to remain unaffected despite an overall decrease in call volumes. Specifically, EMS calls related to traffic and Covid-19 symptoms decreased while calls related to cardiac arrest, stroke, mental health, overdose, and assault did not. These results resonate with those in other geographical areas around the globe. Meanwhile, the response time of EMS was significantly prolonged since the outbreak of the Covid-19 pandemic. This information can provide an opportunity for dialogue among EMS agencies, government officials, and health care partners on adjusting changes in EMS demand and shortening response time during future pandemics.

Moreover, we proposed a time series regression model with change point detection to predict the daily frequency of pandemic EMS incidents. The daily new hospitalization of Covid-19 patients proved a powerful predictor for the number of Covid-19-related EMS calls. In particular, this model suggested that for every 2.5 cases where EMS took a Covid-19 patient to a hospital, one person was admitted. This model may be of interest to other EMS departments as they plan for future pandemics, including the ability to predict pandemic-related calls in an effort to adjust a targeted response. 

\section*{Acknowledgement}
This project is funded by the Good Systems – UT Austin Bridging Barriers Initiative (\url{https://bridgingbarriers.utexas.edu/good-systems/}).
We would like to thank Prof. Junfeng Jiao for suggesting the CDC SVI data and Mauricio Tec for providing data on the hospital admission of Covid-19 patients in the Austin-Round Rock metropolitan statistical area. We also thank Natasha Stewart for the insightful comments on earlier drafts.
\bibliography{mybibfile}

\begin{thebibliography}{77}
\expandafter\ifx\csname natexlab\endcsname\relax\def\natexlab#1{#1}\fi
\providecommand{\url}[1]{\texttt{#1}}
\providecommand{\href}[2]{#2}
\providecommand{\path}[1]{#1}
\providecommand{\DOIprefix}{doi:}
\providecommand{\ArXivprefix}{arXiv:}
\providecommand{\URLprefix}{URL: }
\providecommand{\Pubmedprefix}{pmid:}
\providecommand{\doi}[1]{\href{http://dx.doi.org/#1}{\path{#1}}}
\providecommand{\Pubmed}[1]{\href{pmid:#1}{\path{#1}}}
\providecommand{\bibinfo}[2]{#2}
\ifx\xfnm\relax \def\xfnm[#1]{\unskip,\space#1}\fi
\bibitem[{Organization et~al.(2020)}]{world2020coronavirus}
\bibinfo{author}{W.~H. Organization}, et~al.,
\newblock Coronavirus disease 2019 (COVID-19): situation report, 86
  (\bibinfo{year}{2020}).
\bibitem[{Moghadas et~al.(2020)Moghadas, Shoukat, Fitzpatrick, Wells, Sah,
  Pandey, Sachs, Wang, Meyers, Singer, and Galvani}]{Moghadas9122}
\bibinfo{author}{S.~M. Moghadas}, \bibinfo{author}{A.~Shoukat},
  \bibinfo{author}{M.~C. Fitzpatrick}, \bibinfo{author}{C.~R. Wells},
  \bibinfo{author}{P.~Sah}, \bibinfo{author}{A.~Pandey}, \bibinfo{author}{J.~D.
  Sachs}, \bibinfo{author}{Z.~Wang}, \bibinfo{author}{L.~A. Meyers},
  \bibinfo{author}{B.~H. Singer}, \bibinfo{author}{A.~P. Galvani},
\newblock Projecting hospital utilization during the COVID-19 outbreaks in the
  United States,
\newblock \bibinfo{journal}{Proceedings of the National Academy of Sciences}
  \bibinfo{volume}{117} (\bibinfo{year}{2020}) \bibinfo{pages}{9122--9126}.
\bibitem[{Willan et~al.(2020)Willan, King, Jeffery, and Bienz}]{Willanm1117}
\bibinfo{author}{J.~Willan}, \bibinfo{author}{A.~J. King},
  \bibinfo{author}{K.~Jeffery}, \bibinfo{author}{N.~Bienz},
\newblock Challenges for NHS hospitals during covid-19 epidemic,
\newblock \bibinfo{journal}{BMJ} \bibinfo{volume}{368} (\bibinfo{year}{2020}).
\bibitem[{Guharoy et~al.(2021)Guharoy, Krenzelok, and
  Noviasky}]{guharoy2021race}
\bibinfo{author}{R.~Guharoy}, \bibinfo{author}{E.~Krenzelok},
  \bibinfo{author}{J.~Noviasky},
\newblock Race Against the Pandemic: The United States and Global Health,
\newblock \bibinfo{journal}{Journal of Emergency Medicine} \bibinfo{volume}{60}
  (\bibinfo{year}{2021}) \bibinfo{pages}{402--406}.
\bibitem[{Lerner et~al.(2020)Lerner, Newgard, and Mann}]{LNM20}
\bibinfo{author}{E.~B. Lerner}, \bibinfo{author}{C.~D. Newgard},
  \bibinfo{author}{N.~C. Mann},
\newblock Effect of the Coronavirus Disease 2019 (COVID-19) Pandemic on the
  U.S. Emergency Medical Services System: A Preliminary Report,
\newblock \bibinfo{journal}{Academic Emergency Medicine} \bibinfo{volume}{27}
  (\bibinfo{year}{2020}) \bibinfo{pages}{693--699}.
\bibitem[{Boserup et~al.(2020)Boserup, McKenney, and Elkbuli}]{BOSERUP20201732}
\bibinfo{author}{B.~Boserup}, \bibinfo{author}{M.~McKenney},
  \bibinfo{author}{A.~Elkbuli},
\newblock The impact of the COVID-19 pandemic on emergency department visits
  and patient safety in the United States,
\newblock \bibinfo{journal}{The American Journal of Emergency Medicine}
  \bibinfo{volume}{38} (\bibinfo{year}{2020}) \bibinfo{pages}{1732--1736}.
\bibitem[{Hartnett et~al.(2020)Hartnett, Kite-Powell, DeVies, Coletta, Boehmer,
  Adjemian, Gundlapalli et~al.}]{hartnett2020impact}
\bibinfo{author}{K.~P. Hartnett}, \bibinfo{author}{A.~Kite-Powell},
  \bibinfo{author}{J.~DeVies}, \bibinfo{author}{M.~A. Coletta},
  \bibinfo{author}{T.~K. Boehmer}, \bibinfo{author}{J.~Adjemian},
  \bibinfo{author}{A.~V. Gundlapalli}, et~al.,
\newblock Impact of the COVID-19 pandemic on emergency department
  visits—United States, January 1, 2019--May 30, 2020,
\newblock \bibinfo{journal}{Morbidity and Mortality Weekly Report}
  \bibinfo{volume}{69} (\bibinfo{year}{2020}) \bibinfo{pages}{699}.
\bibitem[{Jeffery et~al.(2020)Jeffery, D’Onofrio, Paek, Platts-Mills, Soares,
  Hoppe, Genes, Nath, and Melnick}]{jeffery2020trends}
\bibinfo{author}{M.~M. Jeffery}, \bibinfo{author}{G.~D’Onofrio},
  \bibinfo{author}{H.~Paek}, \bibinfo{author}{T.~F. Platts-Mills},
  \bibinfo{author}{W.~E. Soares}, \bibinfo{author}{J.~A. Hoppe},
  \bibinfo{author}{N.~Genes}, \bibinfo{author}{B.~Nath}, \bibinfo{author}{E.~R.
  Melnick},
\newblock Trends in emergency department visits and hospital admissions in
  health care systems in 5 states in the first months of the COVID-19 pandemic
  in the US,
\newblock \bibinfo{journal}{JAMA internal medicine} \bibinfo{volume}{180}
  (\bibinfo{year}{2020}) \bibinfo{pages}{1328--1333}.
\bibitem[{Lazzerini et~al.(2020)Lazzerini, Barbi, Apicella, Marchetti,
  Cardinale, and Trobia}]{lazzerini2020delayed}
\bibinfo{author}{M.~Lazzerini}, \bibinfo{author}{E.~Barbi},
  \bibinfo{author}{A.~Apicella}, \bibinfo{author}{F.~Marchetti},
  \bibinfo{author}{F.~Cardinale}, \bibinfo{author}{G.~Trobia},
\newblock Delayed access or provision of care in Italy resulting from fear of
  COVID-19,
\newblock \bibinfo{journal}{The Lancet Child \& Adolescent Health}
  \bibinfo{volume}{4} (\bibinfo{year}{2020}) \bibinfo{pages}{e10--e11}.
\bibitem[{Thornton(2020)}]{thornton2020covid}
\bibinfo{author}{J.~Thornton}, Covid-19: A\&E visits in England fall by 25\% in
  week after lockdown, \bibinfo{year}{2020}.
\bibitem[{Al-Wathinani et~al.(2021)Al-Wathinani, Hertelendy, Alhurishi, Mobrad,
  Alhazmi, Altuwaijri, Alanazi, Alotaibi, and Goniewicz}]{SaudiArabiaCovid}
\bibinfo{author}{A.~Al-Wathinani}, \bibinfo{author}{A.~J. Hertelendy},
  \bibinfo{author}{S.~Alhurishi}, \bibinfo{author}{A.~Mobrad},
  \bibinfo{author}{R.~Alhazmi}, \bibinfo{author}{M.~Altuwaijri},
  \bibinfo{author}{M.~Alanazi}, \bibinfo{author}{R.~Alotaibi},
  \bibinfo{author}{K.~Goniewicz},
\newblock Increased Emergency Calls during the COVID-19 Pandemic in Saudi
  Arabia: A National Retrospective Study,
\newblock \bibinfo{journal}{Healthcare} \bibinfo{volume}{9}
  (\bibinfo{year}{2021}).
\bibitem[{{\c{S}}an et~al.(2020){\c{S}}an, Usul, Bekg{\"o}z, and
  Korkut}]{csan2020effects}
\bibinfo{author}{{\.I}.~{\c{S}}an}, \bibinfo{author}{E.~Usul},
  \bibinfo{author}{B.~Bekg{\"o}z}, \bibinfo{author}{S.~Korkut},
\newblock Effects of COVID-19 pandemic on emergency medical services,
\newblock \bibinfo{journal}{International Journal of Clinical Practice}
  (\bibinfo{year}{2020}) \bibinfo{pages}{e13885}.
\bibitem[{WongLaura et~al.(2020)WongLaura, HawkinsJessica, MurrellKaren
  et~al.}]{wonglaura2020all}
\bibinfo{author}{E.~WongLaura}, \bibinfo{author}{E.~HawkinsJessica},
  \bibinfo{author}{L.~MurrellKaren}, et~al.,
\newblock Where are all the patients? Addressing Covid-19 fear to encourage
  sick patients to seek emergency care,
\newblock \bibinfo{journal}{NEJM Catalyst Innovations in Care Delivery}
  (\bibinfo{year}{2020}).
\bibitem[{Jensen et~al.(2020)Jensen, Holgersen, Jespersen, Blomberg, Folke,
  Lippert, and Christensen}]{jensen2020strategies}
\bibinfo{author}{T.~Jensen}, \bibinfo{author}{M.~G. Holgersen},
  \bibinfo{author}{M.~S. Jespersen}, \bibinfo{author}{S.~N. Blomberg},
  \bibinfo{author}{F.~Folke}, \bibinfo{author}{F.~Lippert},
  \bibinfo{author}{H.~C. Christensen},
\newblock Strategies to handle increased demand in the COVID-19 crisis: A
  coronavirus EMS support track and a web-based self-triage system,
\newblock \bibinfo{journal}{Prehospital Emergency Care} \bibinfo{volume}{25}
  (\bibinfo{year}{2020}) \bibinfo{pages}{28--38}.
\bibitem[{Lucero et~al.(2020)Lucero, Lee, Hyun, Lee, Kahwaji, Miller, Neeki,
  Tamayo-Sarver, and Pan}]{lucero2020underutilization}
\bibinfo{author}{A.~D. Lucero}, \bibinfo{author}{A.~Lee},
  \bibinfo{author}{J.~Hyun}, \bibinfo{author}{C.~Lee},
  \bibinfo{author}{C.~Kahwaji}, \bibinfo{author}{G.~Miller},
  \bibinfo{author}{M.~Neeki}, \bibinfo{author}{J.~Tamayo-Sarver},
  \bibinfo{author}{L.~Pan},
\newblock Underutilization of the Emergency Department During the COVID-19
  Pandemic,
\newblock \bibinfo{journal}{Western Journal of Emergency Medicine}
  \bibinfo{volume}{21} (\bibinfo{year}{2020}) \bibinfo{pages}{15}.
\bibitem[{Butt et~al.(2020)Butt, Azad, Kartha, Masoodi, Bertollini, and
  Abou-Samra}]{butt2020volume}
\bibinfo{author}{A.~A. Butt}, \bibinfo{author}{A.~M. Azad},
  \bibinfo{author}{A.~B. Kartha}, \bibinfo{author}{N.~A. Masoodi},
  \bibinfo{author}{R.~Bertollini}, \bibinfo{author}{A.-B. Abou-Samra},
\newblock Volume and acuity of emergency department visits prior to and after
  COVID-19,
\newblock \bibinfo{journal}{The Journal of emergency medicine}
  \bibinfo{volume}{59} (\bibinfo{year}{2020}) \bibinfo{pages}{730--734}.
\bibitem[{Montagnon et~al.(2021)Montagnon, Rouffilange, Agard, Benner, Cazes,
  and Renard}]{montagnon2021impact}
\bibinfo{author}{R.~Montagnon}, \bibinfo{author}{L.~Rouffilange},
  \bibinfo{author}{G.~Agard}, \bibinfo{author}{P.~Benner},
  \bibinfo{author}{N.~Cazes}, \bibinfo{author}{A.~Renard},
\newblock Impact of the COVID-19 pandemic on emergency department use: focus on
  patients requiring urgent revascularization,
\newblock \bibinfo{journal}{The Journal of Emergency Medicine}
  \bibinfo{volume}{60} (\bibinfo{year}{2021}) \bibinfo{pages}{229--236}.
\bibitem[{Fernandez et~al.(2020)Fernandez, Crowe, Bourn, Matt, Brown, Hawthorn,
  and Brent~Myers}]{fernandez2020covid}
\bibinfo{author}{A.~R. Fernandez}, \bibinfo{author}{R.~P. Crowe},
  \bibinfo{author}{S.~Bourn}, \bibinfo{author}{S.~E. Matt},
  \bibinfo{author}{A.~L. Brown}, \bibinfo{author}{A.~B. Hawthorn},
  \bibinfo{author}{J.~Brent~Myers},
\newblock COVID-19 preliminary case series: characteristics of EMS encounters
  with linked hospital diagnoses,
\newblock \bibinfo{journal}{Prehospital Emergency Care} \bibinfo{volume}{25}
  (\bibinfo{year}{2020}) \bibinfo{pages}{16--27}.
\bibitem[{Yang et~al.(2020)Yang, Barnard, Emert, Drucker, Schwarcz, Counts,
  Murphy, Guan, Kume, Rodriquez et~al.}]{yang2020clinical}
\bibinfo{author}{B.~Y. Yang}, \bibinfo{author}{L.~M. Barnard},
  \bibinfo{author}{J.~M. Emert}, \bibinfo{author}{C.~Drucker},
  \bibinfo{author}{L.~Schwarcz}, \bibinfo{author}{C.~R. Counts},
  \bibinfo{author}{D.~L. Murphy}, \bibinfo{author}{S.~Guan},
  \bibinfo{author}{K.~Kume}, \bibinfo{author}{K.~Rodriquez}, et~al.,
\newblock Clinical characteristics of patients with coronavirus disease 2019
  (COVID-19) receiving emergency medical services in King County, Washington,
\newblock \bibinfo{journal}{JAMA network open} \bibinfo{volume}{3}
  (\bibinfo{year}{2020}) \bibinfo{pages}{e2014549--e2014549}.
\bibitem[{Gibson et~al.(2020)Gibson, Ventura, and Collier}]{GIBSON2020resource}
\bibinfo{author}{C.~Gibson}, \bibinfo{author}{C.~Ventura},
  \bibinfo{author}{G.~D. Collier},
\newblock Emergency Medical Services resource capacity and competency amid
  COVID-19 in the United States: preliminary findings from a national survey,
\newblock \bibinfo{journal}{Heliyon} \bibinfo{volume}{6} (\bibinfo{year}{2020})
  \bibinfo{pages}{e03900}.
\bibitem[{Hick et~al.(2020)Hick, Hanfling, Wynia, and Pavia}]{hick2020duty}
\bibinfo{author}{J.~L. Hick}, \bibinfo{author}{D.~Hanfling},
  \bibinfo{author}{M.~K. Wynia}, \bibinfo{author}{A.~T. Pavia},
\newblock Duty to plan: health care, crisis standards of care, and novel
  coronavirus SARS-CoV-2,
\newblock \bibinfo{journal}{NAM Perspectives}  (\bibinfo{year}{2020}).
\bibitem[{Devereaux et~al.(2020)Devereaux, Yang, Seda, Sankar, Maves, Karanjia,
  Parrish, Rosenberg, Goodman-Crews, Cederquist, and
  et~al.}]{devereaux2020optimizing}
\bibinfo{author}{A.~Devereaux}, \bibinfo{author}{H.~Yang},
  \bibinfo{author}{G.~Seda}, \bibinfo{author}{V.~Sankar},
  \bibinfo{author}{R.~C. Maves}, \bibinfo{author}{N.~Karanjia},
  \bibinfo{author}{J.~S. Parrish}, \bibinfo{author}{C.~Rosenberg},
  \bibinfo{author}{P.~Goodman-Crews}, \bibinfo{author}{L.~Cederquist},
  \bibinfo{author}{et~al.},
\newblock Optimizing Scarce Resource Allocation During COVID-19: Rapid Creation
  of a Regional Health-Care Coalition and Triage Teams in San Diego County,
  California,
\newblock \bibinfo{journal}{Disaster Medicine and Public Health Preparedness}
  (\bibinfo{year}{2020}) \bibinfo{pages}{1–7}.
\bibitem[{Schreyer et~al.(2020)Schreyer, Daniel, King, Blome, DeAngelis,
  Stauffer, Desrochers, Donahue, Politarhos, Raab
  et~al.}]{schreyer2020emergency}
\bibinfo{author}{K.~E. Schreyer}, \bibinfo{author}{A.~Daniel},
  \bibinfo{author}{L.~L. King}, \bibinfo{author}{A.~Blome},
  \bibinfo{author}{M.~DeAngelis}, \bibinfo{author}{K.~Stauffer},
  \bibinfo{author}{K.~Desrochers}, \bibinfo{author}{W.~Donahue},
  \bibinfo{author}{N.~Politarhos}, \bibinfo{author}{C.~Raab}, et~al.,
\newblock Emergency department management of the Covid-19 pandemic,
\newblock \bibinfo{journal}{The Journal of emergency medicine}
  \bibinfo{volume}{59} (\bibinfo{year}{2020}) \bibinfo{pages}{946--951}.
\bibitem[{Murphy et~al.(2020)Murphy, Barnard, Drucker, Yang, Emert, Schwarcz,
  Counts, Jacinto, McCoy, Morgan, Whitney, Bodenman, Duchin, Sayre, and
  Rea}]{Murphy707}
\bibinfo{author}{D.~L. Murphy}, \bibinfo{author}{L.~M. Barnard},
  \bibinfo{author}{C.~J. Drucker}, \bibinfo{author}{B.~Y. Yang},
  \bibinfo{author}{J.~M. Emert}, \bibinfo{author}{L.~Schwarcz},
  \bibinfo{author}{C.~R. Counts}, \bibinfo{author}{T.~Y. Jacinto},
  \bibinfo{author}{A.~M. McCoy}, \bibinfo{author}{T.~A. Morgan},
  \bibinfo{author}{J.~E. Whitney}, \bibinfo{author}{J.~V. Bodenman},
  \bibinfo{author}{J.~S. Duchin}, \bibinfo{author}{M.~R. Sayre},
  \bibinfo{author}{T.~D. Rea},
\newblock Occupational exposures and programmatic response to COVID-19
  pandemic: an emergency medical services experience,
\newblock \bibinfo{journal}{Emergency Medicine Journal} \bibinfo{volume}{37}
  (\bibinfo{year}{2020}) \bibinfo{pages}{707--713}.
\bibitem[{Jalili(2020)}]{jalili2020should}
\bibinfo{author}{M.~Jalili},
\newblock How should emergency medical services personnel protect themselves
  and the patients during COVID-19 pandemic?,
\newblock \bibinfo{journal}{Frontiers in Emergency Medicine}
  \bibinfo{volume}{4} (\bibinfo{year}{2020}) \bibinfo{pages}{e37--e37}.
\bibitem[{Ehrlich et~al.(2020)Ehrlich, McKenney, and
  Elkbuli}]{ehrlich2020defending}
\bibinfo{author}{H.~Ehrlich}, \bibinfo{author}{M.~McKenney},
  \bibinfo{author}{A.~Elkbuli},
\newblock Defending the front lines during the COVID-19 pandemic: Protecting
  our first responders and emergency medical service personnel,
\newblock \bibinfo{journal}{The American Journal of Emergency Medicine}
  (\bibinfo{year}{2020}).
\bibitem[{Spina et~al.(2020)Spina, Marrazzo, Migliari, Stucchi, Sforza, and
  Fumagalli}]{spina2020response}
\bibinfo{author}{S.~Spina}, \bibinfo{author}{F.~Marrazzo},
  \bibinfo{author}{M.~Migliari}, \bibinfo{author}{R.~Stucchi},
  \bibinfo{author}{A.~Sforza}, \bibinfo{author}{R.~Fumagalli},
\newblock The response of Milan's Emergency Medical System to the COVID-19
  outbreak in Italy,
\newblock \bibinfo{journal}{The Lancet} \bibinfo{volume}{395}
  (\bibinfo{year}{2020}) \bibinfo{pages}{e49--e50}.
\bibitem[{Caba{\~n}as et~al.(2020)Caba{\~n}as, Williams, Gallagher, and
  Brice}]{cabanas2020covid}
\bibinfo{author}{J.~G. Caba{\~n}as}, \bibinfo{author}{J.~G. Williams},
  \bibinfo{author}{J.~M. Gallagher}, \bibinfo{author}{J.~H. Brice},
\newblock COVID-19 Pandemic: The Role of EMS Physicians in a Community Response
  Effort,
\newblock \bibinfo{journal}{Prehospital Emergency Care}  (\bibinfo{year}{2020})
  \bibinfo{pages}{1--8}.
\bibitem[{Ghazali et~al.(2020)Ghazali, Ouersighni, Gay, Audebault, Pavlovsky,
  and Casalino}]{gogapc2020Manage}
\bibinfo{author}{D.~A. Ghazali}, \bibinfo{author}{A.~Ouersighni},
  \bibinfo{author}{M.~Gay}, \bibinfo{author}{V.~Audebault},
  \bibinfo{author}{T.~Pavlovsky}, \bibinfo{author}{E.~Casalino},
\newblock Feedback to Prepare EMS Teams to Manage Infected Patients with
  COVID-19: A Case Series,
\newblock \bibinfo{journal}{Prehospital and Disaster Medicine}
  \bibinfo{volume}{35} (\bibinfo{year}{2020}) \bibinfo{pages}{451–453}.
\bibitem[{Maguire et~al.(2020)Maguire, Shearer, McKeown, Phelps, Gerard,
  Handal, and O’Neill}]{maguire2020ethics}
\bibinfo{author}{B.~J. Maguire}, \bibinfo{author}{K.~Shearer},
  \bibinfo{author}{J.~McKeown}, \bibinfo{author}{S.~Phelps},
  \bibinfo{author}{D.~R. Gerard}, \bibinfo{author}{K.~A. Handal},
  \bibinfo{author}{B.~O’Neill},
\newblock The ethics of PPE and EMS in the COVID-19 era,
\newblock \bibinfo{journal}{JEMS}  (\bibinfo{year}{2020}).
\bibitem[{Shadyab et~al.(2021)Shadyab, Castillo, Brennan, Chan, and
  Tolia}]{shadyab2021ethnic}
\bibinfo{author}{A.~H. Shadyab}, \bibinfo{author}{E.~M. Castillo},
  \bibinfo{author}{J.~J. Brennan}, \bibinfo{author}{T.~C. Chan},
  \bibinfo{author}{V.~M. Tolia},
\newblock Ethnic Disparities in COVID-19 among Older Adults Presenting to the
  Geriatric Emergency Department,
\newblock \bibinfo{journal}{The Journal of Emergency Medicine}
  (\bibinfo{year}{2021}).
\bibitem[{Penverne et~al.(2020)Penverne, Jenvrin, and
  Montassier}]{penverne2020ems}
\bibinfo{author}{Y.~Penverne}, \bibinfo{author}{J.~Jenvrin},
  \bibinfo{author}{E.~Montassier},
\newblock EMS dispatch center activity during the COVID-19 containment,
\newblock \bibinfo{journal}{The American Journal of Emergency Medicine}
  (\bibinfo{year}{2020}).
\bibitem[{Satty et~al.(2020)Satty, Ramgopal, Elmer, Mosesso, and
  Martin-Gill}]{satty2020ems}
\bibinfo{author}{T.~Satty}, \bibinfo{author}{S.~Ramgopal},
  \bibinfo{author}{J.~Elmer}, \bibinfo{author}{V.~N. Mosesso},
  \bibinfo{author}{C.~Martin-Gill},
\newblock EMS responses and non-transports during the COVID-19 pandemic,
\newblock \bibinfo{journal}{The American Journal of Emergency Medicine}
  (\bibinfo{year}{2020}).
\bibitem[{Ferron et~al.(2021)Ferron, Agarwal, Cooper, and
  Munkley}]{Ferron21effect}
\bibinfo{author}{R.~Ferron}, \bibinfo{author}{G.~Agarwal},
  \bibinfo{author}{R.~Cooper}, \bibinfo{author}{D.~Munkley},
\newblock The effect of COVID-19 on emergency medical service call volumes and
  patient acuity: a cross-sectional study in Niagara, Ontario,
\newblock \bibinfo{journal}{BMC Emergency Medicine} \bibinfo{volume}{21}
  (\bibinfo{year}{2021}) \bibinfo{pages}{39}.
\bibitem[{Mantica et~al.(2020)Mantica, Riccardi, Terrone, and
  Gratarola}]{mantica2020non}
\bibinfo{author}{G.~Mantica}, \bibinfo{author}{N.~Riccardi},
  \bibinfo{author}{C.~Terrone}, \bibinfo{author}{A.~Gratarola},
\newblock Non-COVID-19 visits to emergency departments during the pandemic: the
  impact of fear,
\newblock \bibinfo{journal}{Public Health} \bibinfo{volume}{183}
  (\bibinfo{year}{2020}) \bibinfo{pages}{40}.
\bibitem[{Laukkanen et~al.(2021)Laukkanen, Lahtinen, Liisanantti, Kaakinen,
  Ehrola, and Raatiniemi}]{Laukkanen21Early}
\bibinfo{author}{L.~Laukkanen}, \bibinfo{author}{S.~Lahtinen},
  \bibinfo{author}{J.~Liisanantti}, \bibinfo{author}{T.~Kaakinen},
  \bibinfo{author}{A.~Ehrola}, \bibinfo{author}{L.~Raatiniemi},
\newblock {Early impact of the COVID-19 pandemic and social restrictions on
  ambulance missions},
\newblock \bibinfo{journal}{European Journal of Public Health}
  (\bibinfo{year}{2021}). \bibinfo{note}{Ckab065}.
\bibitem[{Hall(1971)}]{hall1971management}
\bibinfo{author}{W.~K. Hall},
\newblock Management science approaches to the determination of urban ambulance
  requirements,
\newblock \bibinfo{journal}{Socio-economic Planning Sciences}
  \bibinfo{volume}{5} (\bibinfo{year}{1971}) \bibinfo{pages}{491--499}.
\bibitem[{Aldrich et~al.(1971)Aldrich, Hisserich, and
  Lave}]{aldrich1971analysis}
\bibinfo{author}{C.~A. Aldrich}, \bibinfo{author}{J.~C. Hisserich},
  \bibinfo{author}{L.~B. Lave},
\newblock An analysis of the demand for emergency ambulance service in an urban
  area.,
\newblock \bibinfo{journal}{American Journal of Public Health}
  \bibinfo{volume}{61} (\bibinfo{year}{1971}) \bibinfo{pages}{1156--1169}.
\bibitem[{Siler(1975)}]{siler1975predicting}
\bibinfo{author}{K.~F. Siler},
\newblock Predicting demand for publicly dispatched ambulances in a
  metropolitan area.,
\newblock \bibinfo{journal}{Health Services Research} \bibinfo{volume}{10}
  (\bibinfo{year}{1975}) \bibinfo{pages}{254}.
\bibitem[{Kv{\aa}lseth and Deems(1979)}]{kvaalseth1979statistical}
\bibinfo{author}{T.~O. Kv{\aa}lseth}, \bibinfo{author}{J.~M. Deems},
\newblock Statistical models of the demand for emergency medical services in an
  urban area.,
\newblock \bibinfo{journal}{American Journal of Public Health}
  \bibinfo{volume}{69} (\bibinfo{year}{1979}) \bibinfo{pages}{250--255}.
\bibitem[{McConnel and Wilson(1998)}]{mcconnel1998demand}
\bibinfo{author}{C.~E. McConnel}, \bibinfo{author}{R.~W. Wilson},
\newblock The demand for prehospital emergency services in an aging society,
\newblock \bibinfo{journal}{Social Science \& Medicine} \bibinfo{volume}{46}
  (\bibinfo{year}{1998}) \bibinfo{pages}{1027--1031}.
\bibitem[{Wong and Lai(2012)}]{Wong60}
\bibinfo{author}{H.~T. Wong}, \bibinfo{author}{P.~C. Lai},
\newblock Weather inference and daily demand for emergency ambulance services,
\newblock \bibinfo{journal}{Emergency Medicine Journal} \bibinfo{volume}{29}
  (\bibinfo{year}{2012}) \bibinfo{pages}{60--64}.
\bibitem[{McCarthy et~al.(2008)McCarthy, Zeger, Ding, Aronsky, Hoot, and
  Kelen}]{mccarthy2008challenge}
\bibinfo{author}{M.~L. McCarthy}, \bibinfo{author}{S.~L. Zeger},
  \bibinfo{author}{R.~Ding}, \bibinfo{author}{D.~Aronsky},
  \bibinfo{author}{N.~R. Hoot}, \bibinfo{author}{G.~D. Kelen},
\newblock The challenge of predicting demand for emergency department services,
\newblock \bibinfo{journal}{Academic Emergency Medicine} \bibinfo{volume}{15}
  (\bibinfo{year}{2008}) \bibinfo{pages}{337--346}.
\bibitem[{Channouf et~al.(2007)Channouf, L’Ecuyer, Ingolfsson, and
  Avramidis}]{channouf2007application}
\bibinfo{author}{N.~Channouf}, \bibinfo{author}{P.~L’Ecuyer},
  \bibinfo{author}{A.~Ingolfsson}, \bibinfo{author}{A.~N. Avramidis},
\newblock The application of forecasting techniques to modeling emergency
  medical system calls in Calgary, Alberta,
\newblock \bibinfo{journal}{Health care management science}
  \bibinfo{volume}{10} (\bibinfo{year}{2007}) \bibinfo{pages}{25--45}.
\bibitem[{Setzler et~al.(2009)Setzler, Saydam, and Park}]{setzler2009ems}
\bibinfo{author}{H.~Setzler}, \bibinfo{author}{C.~Saydam},
  \bibinfo{author}{S.~Park},
\newblock EMS call volume predictions: A comparative study,
\newblock \bibinfo{journal}{Computers \& Operations Research}
  \bibinfo{volume}{36} (\bibinfo{year}{2009}) \bibinfo{pages}{1843--1851}.
\bibitem[{Andrew et~al.(2017)Andrew, Nehme, Bernard, Abramson, Newbigin, Piper,
  Dunlop, Holman, and Smith}]{andrew2017stormy}
\bibinfo{author}{E.~Andrew}, \bibinfo{author}{Z.~Nehme},
  \bibinfo{author}{S.~Bernard}, \bibinfo{author}{M.~J. Abramson},
  \bibinfo{author}{E.~Newbigin}, \bibinfo{author}{B.~Piper},
  \bibinfo{author}{J.~Dunlop}, \bibinfo{author}{P.~Holman},
  \bibinfo{author}{K.~Smith},
\newblock Stormy weather: a retrospective analysis of demand for emergency
  medical services during epidemic thunderstorm asthma,
\newblock \bibinfo{journal}{bmj} \bibinfo{volume}{359} (\bibinfo{year}{2017}).
\bibitem[{Group(2020)}]{covid2020early}
\bibinfo{author}{C.-. A.-U.-I.-I. Group},
\newblock Early indicators of intensive care unit bed requirement during the
  COVID-19 epidemic: A retrospective study in Ile-de-France region, France,
\newblock \bibinfo{journal}{PloS one} \bibinfo{volume}{15}
  (\bibinfo{year}{2020}) \bibinfo{pages}{e0241406}.
\bibitem[{Riou et~al.(2020)}]{riou2020emergency}
\bibinfo{author}{B.~Riou}, et~al.,
\newblock Emergency calls are early indicators of ICU bed requirement during
  the COVID-19 epidemic,
\newblock \bibinfo{journal}{medRxiv}  (\bibinfo{year}{2020}).
\bibitem[{Jaffe et~al.(2020)Jaffe, Sonkin, Strugo, and Zerath}]{JSSZ20}
\bibinfo{author}{E.~Jaffe}, \bibinfo{author}{R.~Sonkin},
  \bibinfo{author}{R.~Strugo}, \bibinfo{author}{E.~Zerath},
\newblock Evolution of emergency medical calls during a pandemic - An emergency
  medical service during the COVID-19 outbreak,
\newblock \bibinfo{journal}{The American journal of emergency medicine}
  (\bibinfo{year}{2020}) \bibinfo{pages}{S0735--6757(20)30527--1}.
\bibitem[{Kulpanowski(2021)}]{dataset1}
\bibinfo{author}{D.~Kulpanowski}, University of Texas Good Systems and ATCEMS
  incidents data, \bibinfo{year}{2021}. \URLprefix
  \url{https://data.austintexas.gov/Public-Safety/University-of-Texas-Good-Systems-and-ATCEMS-incide/8mwi-dpdp}.
\bibitem[{MPD(2021)}]{MPDS}
Medical Priority Dispatch System EMD Cardset, \bibinfo{year}{Accessed on
  08/08/2021.} \URLprefix \url{https://prioritydispatch.net/emd-cardset/}.
\bibitem[{Kulpanowski(2021)}]{dataset2}
\bibinfo{author}{D.~Kulpanowski}, University of Texas Good Systems and ATCEMS
  Hospitalization time series, \bibinfo{year}{2021}. \URLprefix
  \url{https://data.austintexas.gov/Public-Safety/University-of-Texas-Good-Systems-and-ATCEMS-Hospit/2srm-55iz}.
\bibitem[{Tec et~al.(2020)Tec, Lachmann, Fox, Pasco, Woody, Starling, Dahan,
  Gaither, Scott, and Meyers}]{tecaustin}
\bibinfo{author}{M.~Tec}, \bibinfo{author}{M.~Lachmann}, \bibinfo{author}{S.~J.
  Fox}, \bibinfo{author}{R.~Pasco}, \bibinfo{author}{S.~Woody},
  \bibinfo{author}{J.~Starling}, \bibinfo{author}{M.~Dahan},
  \bibinfo{author}{K.~Gaither}, \bibinfo{author}{J.~Scott},
  \bibinfo{author}{L.~A. Meyers},
\newblock Austin COVID-19 transmission estimates and healthcare projections
  (\bibinfo{year}{2020}).
\bibitem[{Scott and Knott(1974)}]{scott1974cluster}
\bibinfo{author}{A.~J. Scott}, \bibinfo{author}{M.~Knott},
\newblock A cluster analysis method for grouping means in the analysis of
  variance,
\newblock \bibinfo{journal}{Biometrics}  (\bibinfo{year}{1974})
  \bibinfo{pages}{507--512}.
\bibitem[{Killick et~al.(2012)Killick, Fearnhead, and
  Eckley}]{killick2012optimal}
\bibinfo{author}{R.~Killick}, \bibinfo{author}{P.~Fearnhead},
  \bibinfo{author}{I.~A. Eckley},
\newblock Optimal detection of changepoints with a linear computational cost,
\newblock \bibinfo{journal}{Journal of the American Statistical Association}
  \bibinfo{volume}{107} (\bibinfo{year}{2012}) \bibinfo{pages}{1590--1598}.
\bibitem[{Wei{\ss}(2008)}]{Wei08}
\bibinfo{author}{C.~H. Wei{\ss}},
\newblock Thinning operations for modeling time series of counts---a survey,
\newblock \bibinfo{journal}{AStA Advances in Statistical Analysis}
  \bibinfo{volume}{92} (\bibinfo{year}{2008}) \bibinfo{pages}{319}.
\bibitem[{Siman-Tov et~al.(2021)Siman-Tov, Strugo, Podolsky, and
  Blushtein}]{siman2021assessment}
\bibinfo{author}{M.~Siman-Tov}, \bibinfo{author}{R.~Strugo},
  \bibinfo{author}{T.~Podolsky}, \bibinfo{author}{O.~Blushtein},
\newblock An assessment of treatment, transport, and refusal incidence in a
  National EMS's routine work during COVID-19,
\newblock \bibinfo{journal}{The American Journal of Emergency Medicine}
  (\bibinfo{year}{2021}).
\bibitem[{Moroni et~al.(2020)Moroni, Gramegna, Ajello, Beneduce, Baldetti,
  Vilca, Cappelletti, Scandroglio, and Azzalini}]{moroni2020collateral}
\bibinfo{author}{F.~Moroni}, \bibinfo{author}{M.~Gramegna},
  \bibinfo{author}{S.~Ajello}, \bibinfo{author}{A.~Beneduce},
  \bibinfo{author}{L.~Baldetti}, \bibinfo{author}{L.~M. Vilca},
  \bibinfo{author}{A.~Cappelletti}, \bibinfo{author}{A.~M. Scandroglio},
  \bibinfo{author}{L.~Azzalini},
\newblock Collateral damage: medical care avoidance behavior among patients
  with myocardial infarction during the COVID-19 pandemic,
\newblock \bibinfo{journal}{Case Reports} \bibinfo{volume}{2}
  (\bibinfo{year}{2020}) \bibinfo{pages}{1620--1624}.
\bibitem[{Doucette et~al.(2021)Doucette, Tucker, Auguste, Watkins, Green,
  Pereira, Borrup, Shapiro, and Lapidus}]{doucette2021initial}
\bibinfo{author}{M.~L. Doucette}, \bibinfo{author}{A.~Tucker},
  \bibinfo{author}{M.~E. Auguste}, \bibinfo{author}{A.~Watkins},
  \bibinfo{author}{C.~Green}, \bibinfo{author}{F.~E. Pereira},
  \bibinfo{author}{K.~T. Borrup}, \bibinfo{author}{D.~Shapiro},
  \bibinfo{author}{G.~Lapidus},
\newblock Initial impact of COVID-19’s stay-at-home order on motor vehicle
  traffic and crash patterns in Connecticut: an interrupted time series
  analysis,
\newblock \bibinfo{journal}{Injury prevention} \bibinfo{volume}{27}
  (\bibinfo{year}{2021}) \bibinfo{pages}{3--9}.
\bibitem[{de~Koning et~al.(2021)de~Koning, Boogers, Bosch, de~Visser, Schalij,
  and Beeres}]{Koning21Emergency}
\bibinfo{author}{E.~R. de~Koning}, \bibinfo{author}{M.~J. Boogers},
  \bibinfo{author}{J.~Bosch}, \bibinfo{author}{M.~de~Visser},
  \bibinfo{author}{M.~J. Schalij}, \bibinfo{author}{S.~L. M.~A. Beeres},
\newblock Emergency medical services evaluations for chest pain during first
  COVID-19 lockdown in Hollands-Midden, the Netherlands,
\newblock \bibinfo{journal}{Netherlands Heart Journal} \bibinfo{volume}{29}
  (\bibinfo{year}{2021}) \bibinfo{pages}{224--229}.
\bibitem[{Goldberg et~al.(2021)Goldberg, Cash, Peters, Weiner, Greenough, and
  Seethala}]{goldberg2021impact}
\bibinfo{author}{S.~A. Goldberg}, \bibinfo{author}{R.~E. Cash},
  \bibinfo{author}{G.~Peters}, \bibinfo{author}{S.~G. Weiner},
  \bibinfo{author}{P.~G. Greenough}, \bibinfo{author}{R.~Seethala},
\newblock The impact of COVID-19 on statewide EMS use for cardiac emergencies
  and stroke in Massachusetts,
\newblock \bibinfo{journal}{Journal of the American College of Emergency
  Physicians Open} \bibinfo{volume}{2} (\bibinfo{year}{2021})
  \bibinfo{pages}{e12351}.
\bibitem[{Muldoon et~al.(2021)Muldoon, Denize, Talarico, Fell, Sobiesiak,
  Heimerl, and Sampsel}]{Muldoon2021violence}
\bibinfo{author}{K.~A. Muldoon}, \bibinfo{author}{K.~M. Denize},
  \bibinfo{author}{R.~Talarico}, \bibinfo{author}{D.~B. Fell},
  \bibinfo{author}{A.~Sobiesiak}, \bibinfo{author}{M.~Heimerl},
  \bibinfo{author}{K.~Sampsel},
\newblock COVID-19 pandemic and violence: rising risks and decreasing urgent
  care-seeking for sexual assault and domestic violence survivors,
\newblock \bibinfo{journal}{BMC Medicine} \bibinfo{volume}{19}
  (\bibinfo{year}{2021}) \bibinfo{pages}{20}.
\bibitem[{Wilde(2013)}]{wilde2013emergency}
\bibinfo{author}{E.~T. Wilde},
\newblock Do emergency medical system response times matter for health
  outcomes?,
\newblock \bibinfo{journal}{Health economics} \bibinfo{volume}{22}
  (\bibinfo{year}{2013}) \bibinfo{pages}{790--806}.
\bibitem[{Blackwell and Kaufman(2002)}]{blackwell2002response}
\bibinfo{author}{T.~H. Blackwell}, \bibinfo{author}{J.~S. Kaufman},
\newblock Response time effectiveness: comparison of response time and survival
  in an urban emergency medical services system,
\newblock \bibinfo{journal}{Academic Emergency Medicine} \bibinfo{volume}{9}
  (\bibinfo{year}{2002}) \bibinfo{pages}{288--295}.
\bibitem[{Pons et~al.(2005)Pons, Haukoos, Bludworth, Cribley, Pons, and
  Markovchick}]{pons2005paramedic}
\bibinfo{author}{P.~T. Pons}, \bibinfo{author}{J.~S. Haukoos},
  \bibinfo{author}{W.~Bludworth}, \bibinfo{author}{T.~Cribley},
  \bibinfo{author}{K.~A. Pons}, \bibinfo{author}{V.~J. Markovchick},
\newblock Paramedic response time: does it affect patient survival?,
\newblock \bibinfo{journal}{Academic Emergency Medicine} \bibinfo{volume}{12}
  (\bibinfo{year}{2005}) \bibinfo{pages}{594--600}.
\bibitem[{for Disease~Control et~al.(2020)for Disease~Control, Prevention
  et~al.}]{centers2020interim}
\bibinfo{author}{C.~for Disease~Control}, \bibinfo{author}{Prevention}, et~al.,
\newblock Interim Recommendations for Emergency Medical Services (EMS) Systems
  and 911 Public Safety Answering Points/Emergency Communication Centers
  (PSAP/ECCS) in the United States During the Coronavirus Disease (COVID-19)
  Pandemic,
\newblock in: \bibinfo{booktitle}{Centers for Disease Control and Prevention
  (US)}, \bibinfo{organization}{Centers for Disease Control and Prevention
  (US)}, \bibinfo{year}{2020}.
\bibitem[{Ribeiro~Jr et~al.(2020)Ribeiro~Jr, Alexandrino, Koleda, Baptista,
  Azfar, Di~Saverio, Ponchietti, G{\"u}emes, Blas, Mesquita
  et~al.}]{ribeiro2020international}
\bibinfo{author}{M.~Ribeiro~Jr}, \bibinfo{author}{H.~Alexandrino},
  \bibinfo{author}{P.~Koleda}, \bibinfo{author}{S.~Baptista},
  \bibinfo{author}{M.~Azfar}, \bibinfo{author}{S.~Di~Saverio},
  \bibinfo{author}{L.~Ponchietti}, \bibinfo{author}{A.~G{\"u}emes},
  \bibinfo{author}{J.~Blas}, \bibinfo{author}{C.~Mesquita}, et~al.,
\newblock International cooperation group of emergency surgery during the
  COVID-19 pandemic.,
\newblock \bibinfo{journal}{European Journal of Trauma and Emergency Surgery:
  Official Publication of the European Trauma Society}  (\bibinfo{year}{2020}).
\bibitem[{Ben{\'I}tez et~al.(2020)Ben{\'I}tez, Pedival, Talal, Cros, Ribeiro,
  Azfar, Saverio, and Laina}]{benitez2020adapting}
\bibinfo{author}{C.~Y. Ben{\'I}tez}, \bibinfo{author}{A.~N. Pedival},
  \bibinfo{author}{I.~Talal}, \bibinfo{author}{B.~Cros},
  \bibinfo{author}{M.~A.~F. Ribeiro}, \bibinfo{author}{M.~Azfar},
  \bibinfo{author}{S.~D. Saverio}, \bibinfo{author}{J.~L.~B. Laina},
\newblock Adapting to an unprecedented scenario: surgery during the COVID-19
  outbreak,
\newblock \bibinfo{journal}{Revista do Col{\'e}gio Brasileiro de
  Cirurgi{\~o}es} \bibinfo{volume}{47} (\bibinfo{year}{2020}).
\bibitem[{Ben{\'\i}tez et~al.(2020)Ben{\'\i}tez, G{\"u}emes, Aranda, Ribeiro,
  Ottolino, Di~Saverio, Alexandrino, Ponchietti, Blas, Ramos
  et~al.}]{benitez2020impact}
\bibinfo{author}{C.~Y. Ben{\'\i}tez}, \bibinfo{author}{A.~G{\"u}emes},
  \bibinfo{author}{J.~Aranda}, \bibinfo{author}{M.~Ribeiro},
  \bibinfo{author}{P.~Ottolino}, \bibinfo{author}{S.~Di~Saverio},
  \bibinfo{author}{H.~Alexandrino}, \bibinfo{author}{L.~Ponchietti},
  \bibinfo{author}{J.~L. Blas}, \bibinfo{author}{J.~P. Ramos}, et~al.,
\newblock Impact of personal protective equipment on surgical performance
  during the COVID-19 pandemic,
\newblock \bibinfo{journal}{World Journal of Surgery} \bibinfo{volume}{44}
  (\bibinfo{year}{2020}) \bibinfo{pages}{2842--2847}.
\bibitem[{Ventura et~al.(2020)Ventura, Gibson, and
  Collier}]{ventura2020emergency}
\bibinfo{author}{C.~Ventura}, \bibinfo{author}{C.~Gibson},
  \bibinfo{author}{G.~D. Collier},
\newblock Emergency Medical Services resource capacity and competency amid
  COVID-19 in the United States: preliminary findings from a national survey,
\newblock \bibinfo{journal}{Heliyon} \bibinfo{volume}{6} (\bibinfo{year}{2020})
  \bibinfo{pages}{e03900}.
\bibitem[{Weber(0723)}]{Weber20Austin}
\bibinfo{author}{A.~Weber}, Austin EMS Medics Are Overworked And Overloaded
  During COVID-19. It's Taking A 'Huge Mental Toll.',
  \bibinfo{year}{2020/07/23}. \URLprefix
  \url{https://www.kut.org/austin/2020-07-23/austin-ems-medics-are-overworked-and-overloaded-during-covid-19-its-taking-a-huge-mental-toll},
  \bibinfo{note}{accessed: 2021/05/21}.
\bibitem[{Balagtas(0722)}]{Tristan20Austin}
\bibinfo{author}{T.~Balagtas}, Austin EMS Association says pandemic has
  impacted emergency response times, \bibinfo{year}{2020/07/22}. \URLprefix
  \url{https://cbsaustin.com/news/local/austin-ems-association-says-pandemic-has-impacted-emergency-response-times},
  \bibinfo{note}{accessed: 2021/05/21}.
\bibitem[{Moghadas et~al.(2020)Moghadas, Shoukat, Fitzpatrick, Wells, Sah,
  Pandey, Sachs, Wang, Meyers, Singer et~al.}]{moghadas2020projecting}
\bibinfo{author}{S.~M. Moghadas}, \bibinfo{author}{A.~Shoukat},
  \bibinfo{author}{M.~C. Fitzpatrick}, \bibinfo{author}{C.~R. Wells},
  \bibinfo{author}{P.~Sah}, \bibinfo{author}{A.~Pandey}, \bibinfo{author}{J.~D.
  Sachs}, \bibinfo{author}{Z.~Wang}, \bibinfo{author}{L.~A. Meyers},
  \bibinfo{author}{B.~H. Singer}, et~al.,
\newblock Projecting hospital utilization during the COVID-19 outbreaks in the
  United States,
\newblock \bibinfo{journal}{Proceedings of the National Academy of Sciences}
  \bibinfo{volume}{117} (\bibinfo{year}{2020}) \bibinfo{pages}{9122--9126}.
\bibitem[{Du et~al.(2020)Du, Wang, Pasco, Petty, Fox, and Meyers}]{du2020covid}
\bibinfo{author}{Z.~Du}, \bibinfo{author}{X.~Wang}, \bibinfo{author}{R.~Pasco},
  \bibinfo{author}{M.~Petty}, \bibinfo{author}{S.~J. Fox},
  \bibinfo{author}{L.~A. Meyers},
\newblock Covid-19 healthcare demand projections: 22 texas cities,
\newblock \bibinfo{journal}{UT COVID-19 Consortium}  (\bibinfo{year}{2020}).
\bibitem[{Gel et~al.(2020)Gel, Jehn, Lant, Muldoon, Nelson, and
  Ross}]{gel2020covid}
\bibinfo{author}{E.~S. Gel}, \bibinfo{author}{M.~Jehn},
  \bibinfo{author}{T.~Lant}, \bibinfo{author}{A.~R. Muldoon},
  \bibinfo{author}{T.~Nelson}, \bibinfo{author}{H.~M. Ross},
\newblock COVID-19 healthcare demand projections: Arizona,
\newblock \bibinfo{journal}{PloS one} \bibinfo{volume}{15}
  (\bibinfo{year}{2020}) \bibinfo{pages}{e0242588}.
\bibitem[{Ferstad et~al.(2020)Ferstad, Gu, Lee, Thapa, Shin, Salomon, Glynn,
  Shah, Milstein, Schulman et~al.}]{ferstad2020model}
\bibinfo{author}{J.~O. Ferstad}, \bibinfo{author}{A.~J. Gu},
  \bibinfo{author}{R.~Y. Lee}, \bibinfo{author}{I.~Thapa},
  \bibinfo{author}{A.~Y. Shin}, \bibinfo{author}{J.~A. Salomon},
  \bibinfo{author}{P.~Glynn}, \bibinfo{author}{N.~H. Shah},
  \bibinfo{author}{A.~Milstein}, \bibinfo{author}{K.~Schulman}, et~al.,
\newblock A model to forecast regional demand for COVID-19 related hospital
  beds,
\newblock \bibinfo{journal}{medRxiv}  (\bibinfo{year}{2020}).
\bibitem[{Chen and Gupta(1997)}]{chen1997testing}
\bibinfo{author}{J.~Chen}, \bibinfo{author}{A.~K. Gupta},
\newblock Testing and locating variance changepoints with application to stock
  prices,
\newblock \bibinfo{journal}{Journal of the American Statistical association}
  \bibinfo{volume}{92} (\bibinfo{year}{1997}) \bibinfo{pages}{739--747}.

\end{thebibliography}

\section*{Appendix}

\subsection*{Comparison of mean response time across hospitals}

\begin{figure}[h]\centering
\caption{Comparison of response time across hospitals - mean plot}
\includegraphics[width=0.7\textwidth]{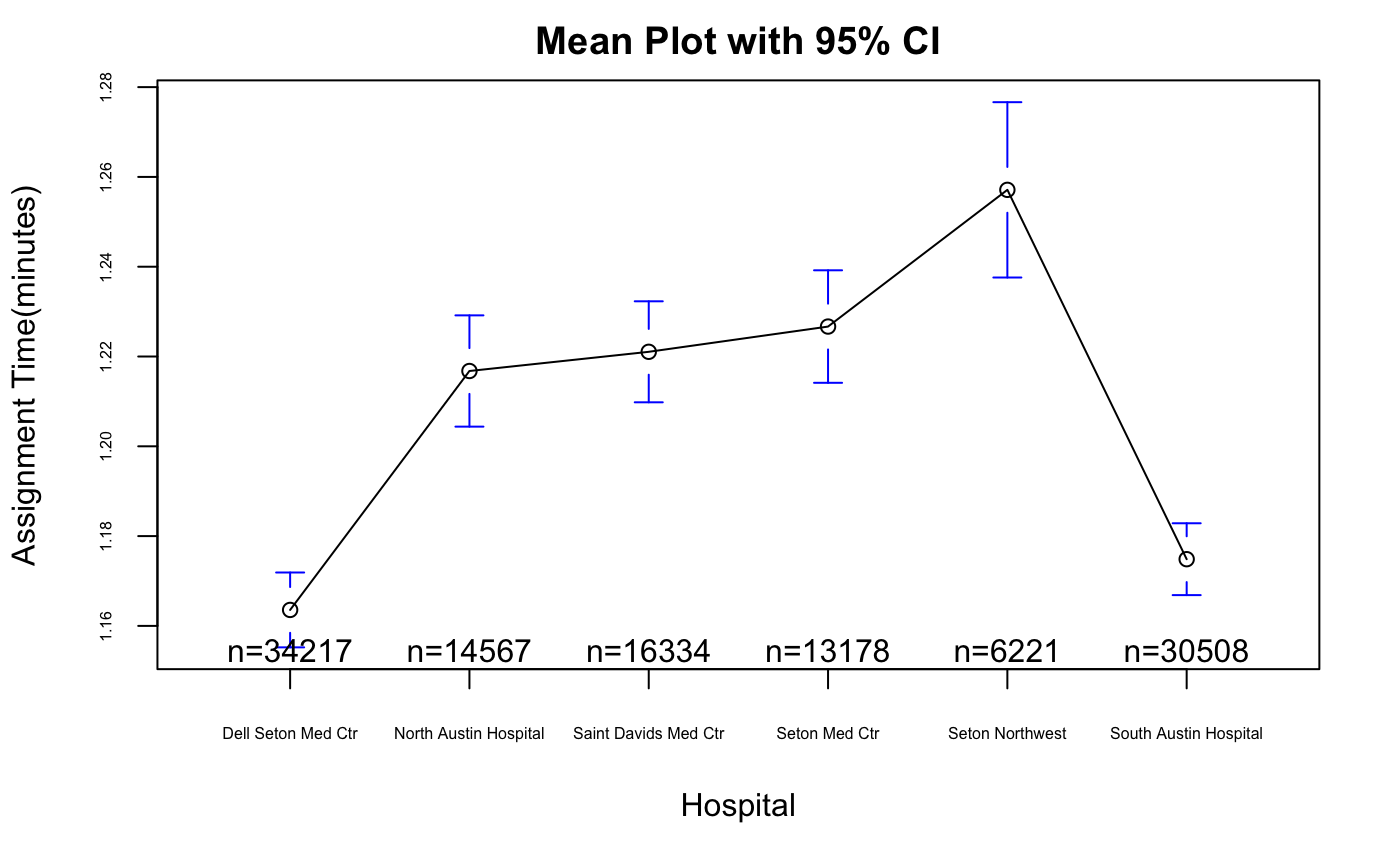}
\includegraphics[width=0.7\textwidth]{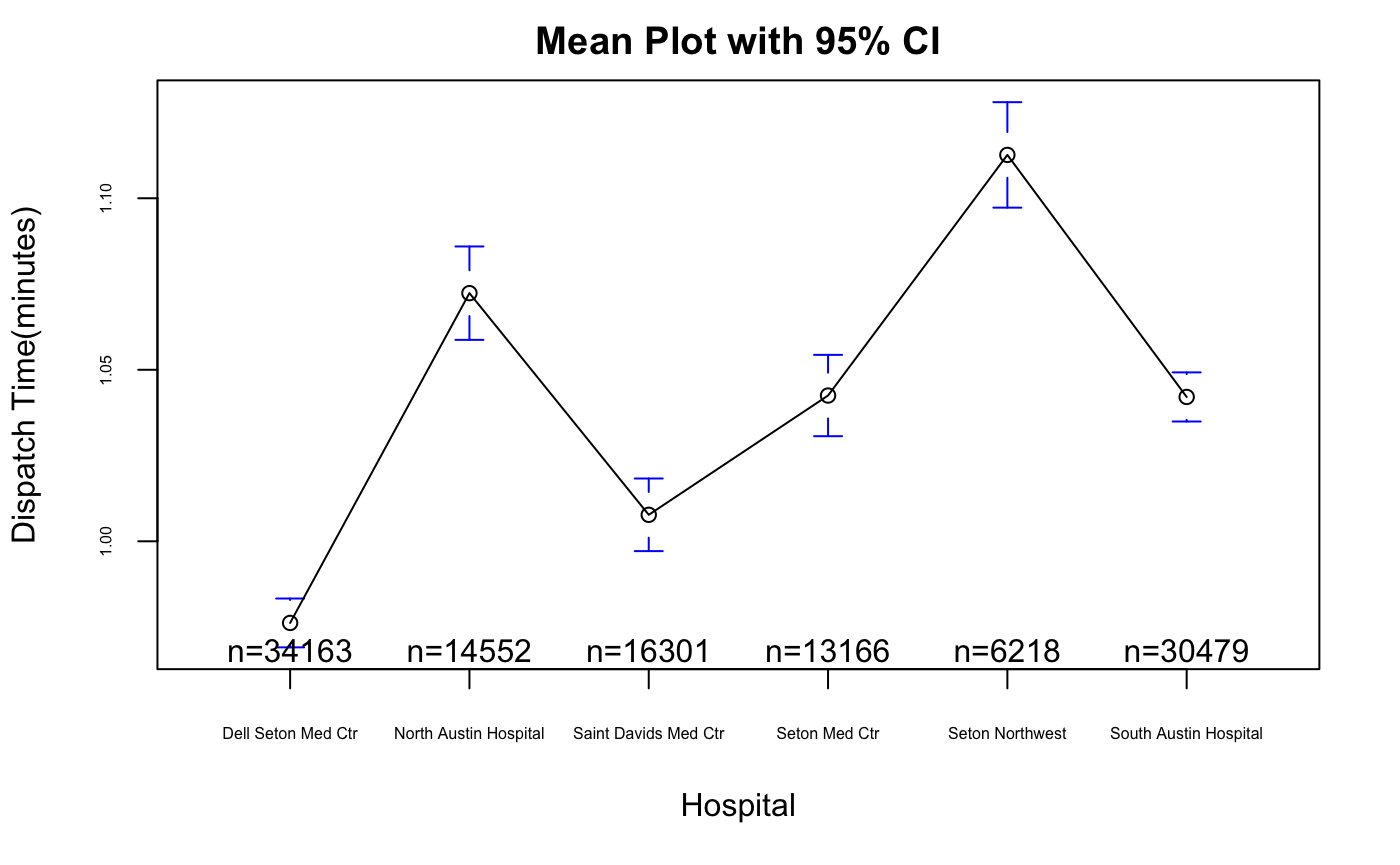}
\includegraphics[width=0.7\textwidth]{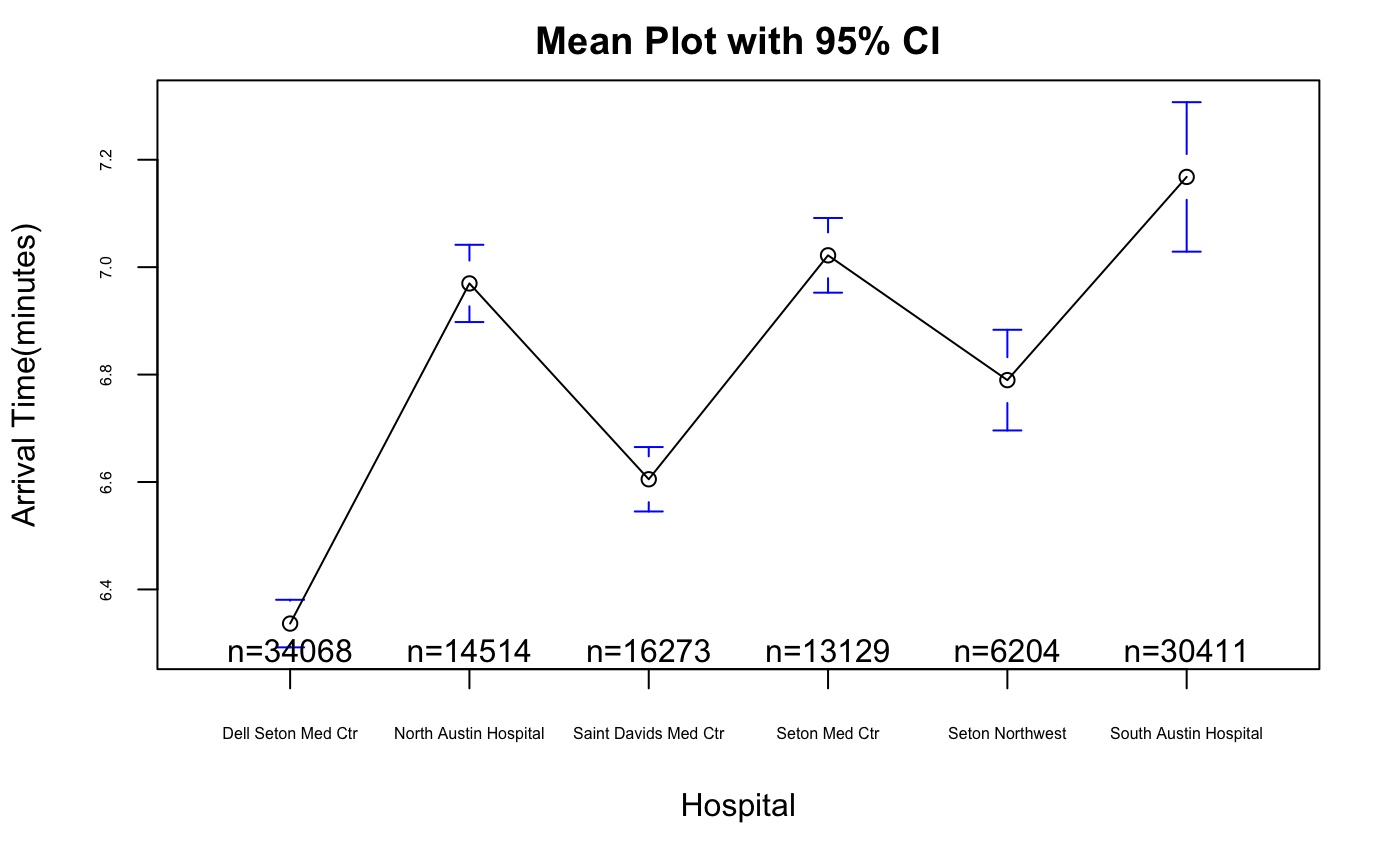}
\label{fig:Comparison Response}
\end{figure}

One-way analysis of variance (ANOVA) suggested that the mean response time varied for the 6 major hospitals (F-test, p-value $< 0.01$ for assignment, dispatch and arrival). This is also shown in the mean plots in Figure \ref{fig:Comparison Response}. In particular, Dell Seton Medical Center, located in downtown Austin, had a lower response time than other hospitals. However, the slight difference in time may not have practical significance. A complete list of the mean and median response time for each of the 6 hospitals in given in Table \ref{table: Comparison Response}. 

\begin{table}[h]
\centering
\begin{tabular}{||c c c c c c c||} 
 \hline
 & \multicolumn{2}{c}{Assignment Time} & \multicolumn{2}{c}{Dispatch Time} & \multicolumn{2}{c||}{Arrival Time}\\ [0.5ex] 
  & Mean & Median & Mean & Median & Mean & Median\\ [0.5ex] 
 \hline\hline
Dell Seton Med Ctr    & 1.16 & 1.03 & 0.98 & 1.00  & 6.34 & 5.57\\
North Austin Hospital	 & 1.22 & 1.08 & 1.07 & 1.08  & 6.97 & 6.20 \\
Saint Davids Med Ctr & 1.22 &	1.08 & 1.01 & 1.03 & 6.61 & 5.88 \\
Seton Med Ctr            & 1.23 &	1.08 & 1.04 & 1.07 & 7.02 & 6.27 \\
Seton Northwest        & 1.26 &	1.12 & 1.11 & 1.13 & 6.79 & 6.07 \\
South Austin Hospital & 1.17 & 1.05 & 1.04 & 1.07 & 7.17 & 6.30 \\
 \hline
\end{tabular}
\caption{Comparison of mean response time across hospitals}
\label{table: Comparison Response}
\end{table}

\subsection*{Comparison of change point locations with respect to various choices of penalty}

To evaluate the impact of pandemic on non-pandemic incidents, we identified changes in mean and variance with binary segmentation method and BIC penalty assuming underlying normal distribution. To restrict our attention to the impact of the pandemic only, we chose the maximal number of change points as 2. When we allowed a greater number of change points, binary segmentation with both BIC penalty and Schwarz information criterion (SIC \cite{chen1997testing}) penalty also identified the date April 8th, around which the non-pandemic incidents began to increase from the lowest point. We have also tested the PELT method, which failed to produce meaningful results. To identify changes in mean and variance, the PELT method was sensitive to random noise and produced too many dates. When we restricted the PELT method to variance only, it failed to identify the date March 17th.

To identify multiple change points in the daily hospitalization data, we applied PELT method on variance with MBIC penalty assuming underlying normal distribution. We chose PELT on variance with MBIC penalty because it produced a relatively small number of change points, which could help avoid overfitting. The PELT method with other types of penalties (BIC, Akaike information criterion (AIC), SIC), as well as the Binary Segmentation method, however, produced too many change points ($\ge 6$). 

\subsection*{Time series regression model without change point detection}

Table \ref{table:coef simple} gives the output of a single variable time series regression model. The mean squared error of this model is 40.501, which is worse than that of our proposed model (table \ref{table:coef}).

\begin{table}[h!]
\centering
\begin{tabular}{||c c c||} 
 \hline
 Coefficients & intercept & hosp\\ [0.5ex] 
 \hline\hline
Estimate & 26.10542 & 0.27774\\ 
Standard Error & 0.94579  & 0.02735\\
 \hline\hline
 train & \multicolumn{2}{c||}{$r^2 = 0.59$} \\
 & \multicolumn{2}{c||}{Residual standard error: 7.665 on 212 degrees of freedom}\\
 test & \multicolumn{2}{c||}{$r^2 = 0.54$} \\
 & \multicolumn{2}{c||}{Mean squared error: 40.501 } \\
 & \multicolumn{2}{c||}{Standard error of prediction residual: 6.420}\\
\hline
\end{tabular}
\caption{Coefficients of Simple Time Series Regression}
\label{table:coef simple}
\end{table}

\end{document}